\documentclass[11pt]{article}
\usepackage{latexsym,axodraw,mathrsfs}

\title{\bf Gravitational decay of the $Z$-boson}
\author{\bf Jos\'e F. Nieves\\
Laboratory of Theoretical Physics\\ 
Department of Physics, P.O. Box 23343\\
University of Puerto Rico, R\'{\i}o Piedras,
Puerto Rico 00931-3343
\and
\bf Palash B. Pal\\
Saha Institute of Nuclear Physics\\ 
1/AF Bidhan-Nagar, Calcutta 700064, India
}
\date{}

\evensidemargin=\oddsidemargin

\makeatletter
\@addtoreset{equation}{section}

\def\Eq#1{Eq.\ (\ref{#1})}
\def\Eqs#1#2{Eqs.\ (\ref{#1}) and (\ref{#2})}
\def\Eqss#1#2#3{Eqs.\ (\ref{#1}), (\ref{#2}) and (\ref{#3})}
\def\fig#1{Fig.~\ref{#1}}

\makeatother

\def\Graviton(#1,#2)(#3,#4)#5#6{\Photon(#1,#2)(#3,#4)#5#6
  \Photon(#1,#2)(#3,#4){-#5}#6}

\def\grav{{\cal G}}

\def\lint#1{\int \! {{\sf d^4} #1 \over (2\pi)^{\sf 4}}}
\def\lintd#1{\int \! {{\sf d^D} #1 \over (2\pi)^{\sf D}}}
\def\slash#1{\rlap/#1}
\def\gtil{\widetilde\gamma}
\def\Ntil{\widetilde N}
\def\Tr{\mathop{\rm Tr}}
\def\lag{\mathscr L}

\def\diagtag#1{  \ifx#10 {} \else {\,:\,{\rm #1}}
  \fi}

\def\di(#1/#2){%
  \ifx#1f {(f\diagtag{#2})} \else
    \ifx#1W {(W\diagtag{#2})} \else
    \fi
  \fi
}

\def\T(#1/#2){%
  \ifx#2N F \else
  \ifx#10 T_{\lambda\rho\mu\nu} \else
  \ifx#11 F_{\lambda\rho\mu\nu} \else
  \ifx#20 T_{\lambda\rho\mu\nu}^{(#1)} \else
  \ifx#21 F_{\lambda\rho\mu\nu}^{(#1)} \else
  T_{\lambda\rho\mu\nu}^{\di(#1/#2)}\fi\fi\fi\fi\fi}

\def\S..{S_{\lambda\rho\mu\nu}}

\newdimen{\notewidth}
\def\ournote((#1)){\setbox0=\hbox{{\bf #1}}
\ifnum\wd0 < \notewidth
\begin{center} \framebox{\Magenta{\bf #1}} \end{center}
\else
\begin{center}\framebox{\parbox{.9\textwidth}{\Magenta{\bf
	#1}}}\end{center}\fi}

\begin{document}

\maketitle

\begin{abstract}
We study the decay process of the $Z$ boson to a photon and a
graviton.  The most general form of the on-shell amplitude, subject to
the constraints due to the conservation of the electromagnetic current
and the energy-momentum tensor, is determined. The amplitude is
expressed in terms of three form factors, two of which are CP-odd
while one is CP-even.  The latter, which is the only non-zero form
factor at the one-loop level, is computed in the standard model and
the decay rate is determined.
\end{abstract}

\section{Introduction}
As is well known \cite{Yang:1950rg}, the $Z$-boson cannot decay into two
photons.  Therefore, the simplest decay of the $Z$-boson to two
massless bosons is through the channel
\begin{eqnarray}
Z(p) \to \gamma(k) + \grav(q) \,,
\end{eqnarray}
where $\gamma$ denotes the photon and $\grav$ the graviton.  In this
work we consider this process and calculate the decay rate, assuming
the standard electroweak interactions and the canonical gravitational
coupling of the standard particles.

We work with the linear theory of gravity, which means
that we write the space-time metric in the form 
\begin{eqnarray}
g_{\lambda\rho} = \eta_{\lambda\rho} + 2\kappa h_{\lambda\rho} 
\label{eta+h}
\end{eqnarray}
where $h_{\lambda\rho}$ is identified with the graviton field,
and the gravitational couplings in the Lagrangian are expanded
up to the linear order in $\kappa$. Furthermore, the constant $\kappa$
is defined in terms of Newton's gravitational constant by
\begin{eqnarray}
\kappa = \sqrt{8\pi G} \,,
\end{eqnarray}
which is such that the field $h_{\lambda\rho}$ has the properly
normalized kinetic energy term in the Lagrangian.  This point of view
for treating processes involving the gravitational and Standard Model
interactions is the same as that employed in some recent works for the
calculation of quantum gravity amplitudes \cite{Bjerrum-Bohr:2002kt,
Bjerrum-Bohr:2002ks}, in which General Relativity is treated as an
effective field theory for energies below the Planck scale
\cite{Weinberg:1978kz, Donoghue:1994dn, Burgess:2003jk}.

The process is a clean one as far the kinematics is concerned: it
involves the photon, with an energy equal to half the $Z$-boson
mass. However, the rate is not sizable, as a simple order-of-magnitude
estimate of the decay rate readily shows.  The gravitational couplings
will involve the factor $\kappa$ in the amplitude, and therefore $G$
in the rate.  The $Z$ and the photon couplings provide a factor
$\alpha^2$ in the rate.  Dimensional arguments then indicate that the
decay rate is of the form
\begin{eqnarray}
\label{estimate}
\Gamma \sim \alpha^2 G M_Z^3 \,,
\end{eqnarray}
which is of order $10^{-36}$\,GeV.

Despite this, we have two motivations for performing this calculation.
First, in theories that extend the standard model of electroweak
interactions and/or the gravitational interactions, including theories
of Lorentz invariance violation, the rate might be different.  This
calculation can pave the way and serve as a test case for similar
calculations in the context of those extended theories.  Second, the
methods employed here can also be helpful in the calculations of
related processes where the $Z$ boson appears as a virtual particle,
such as $\nu\bar\nu\to\gamma\grav$, and it is conceivable that they
can have physical relevance in some astrophysical contexts.

The amplitude for the process is determined by a set of one-loop diagrams
in perturbation theory, that we divide into two classes.
The first class, to which we refer as the fermion loops, consists of
the diagrams that contain fermion lines circulating in the loop.
The second class consists of the diagrams in which the $W$ bosons
circulate in the loops, and in principle the diagrams that contain
their corresponding unphysical Higgs bosons and Fadeev-Popov ghosts.
Among the various vertices required to compute the diagrams,
we need the gravitational couplings of the $W$ boson
as well as those of the $Z$ and the photon. We adopt the point of view
that, for each gauge boson $V = \gamma, W, Z$,
those couplings are determined by the interaction Lagrangian
\begin{equation}
\lag_{\rm int} = -\kappa h^{\lambda\rho} T^{(V)}_{\lambda\rho} \,,
\end{equation}
where $T^{(V)}_{\lambda\rho}$
is the expression for the energy-momentum tensor that is obtained
from the linear expansion of the gravitational Lagrangian term
\begin{equation}
\lag^{(V)}_g = \sqrt{-g}\, \lag^{(V)}_0 \,,
\end{equation}
with $\lag_0$ being the canonical expression for the bilinear part
of the Lagrangian. Thus, for example, for the photon,
\begin{equation}
\label{L0photon}
\lag^{(\gamma)}_0 = -\frac{1}{4}F^{\mu\nu} F_{\mu\nu} \,,
\end{equation}
while for the $W$ and $Z$ it contains the corresponding mass term.
This amounts to adopting the unitary gauge for the $W$ and $Z$ gauge
bosons and therefore, for consistency, we employ the unitary gauge 
throughout.  The amplitude that is obtained in this way has the
following properties, which justify this approach: (1) it satisfies
the transversality condition required by the conservation of the
electromagnetic current, (2) it satisfies the analogous condition
required by the conservation of the energy-momentum tensor, (3) the
diagrams yield a finite contribution to the amplitude. In the
presentation that follows, we consider the fermion loops first which
are simpler, and therefore they allow us to introduce a set of
techniques that are useful for treating the more complicated $W$
loops.

The paper is organized as follows.  In Sec.~\ref{s:gf}, we show that
Lorentz invariance together with the electromagnetic and gravitational
transversality conditions imply that the on-shell amplitude can be
expressed in terms of three form factors, two of which are non-zero
only if the CP symmetry is broken at some level. The remainder of the
paper is devoted to the calculation of the only form factor that is
non-zero at the one-loop level.  In Sec.~\ref{s:fl}, we enumerate the
fermion loop diagrams and the corresponding Feynman rules, and we
verify that they form a gauge invariant set in the sense that their
total contribution to the amplitude satisfies the electromagnetic and
gravitational transversality conditions.  We then proceed to calculate
their contribution to the form factors and, as expected, the two
CP-odd form factors vanish and only the one that is CP even survives
and it is finite.  In Sec.~\ref{s:Wl} we carry out a similar procedure
for the $W$ loop diagrams, with analogous results. As already
mentioned, the calculations are carried out employing the unitary
gauge for the $W,Z$ propagators, together with the gravitational
couplings that follow from the canonical expressions for the
energy-momentum tensor.  Finally, in Sec.~\ref{s:dk}, we use the
results obtained in the previous sections to compute the decay rate.
In the course of the calculations we have used several Ward-like
identities that relate the various gravitational vertices and other
algebraic manipulations, we which have summarized in the appendices.

\section{General form of the amplitude}\label{s:gf}
We introduce the off-shell vertex function $\T(1/)(q, k)$, which is
defined such that the on-shell amplitude for the process is given by
\begin{eqnarray}
\mathscr M =  {\cal E}^{\lambda\rho\ast}(q) \varepsilon^{\nu\ast}(k)
\varepsilon^\mu_Z (p) \T(1/)(q, k) \,,
\label{M}
\end{eqnarray}
where ${\cal E}^{\lambda\rho}$ is the polarization tensor of the
graviton, while $\varepsilon^\nu$ and $\varepsilon^\mu_Z$ are the
polarization vectors for the photon and $Z$, respectively.  Since we
are calculating the on-shell amplitude, the momentum vectors satisfy
the on-shell conditions
\begin{eqnarray}
k^2 &=& 0 \,, 
\label{k^2=0} \\ 
q^2 &=& 0 \,, 
\label{q^2=0} \\ 
p^2 &=& M_Z^2 \,,
\label{p^2}
\end{eqnarray}
and they are related by momentum conservation
\begin{equation}
\label{p=k+q}
p = k + q \,,
\end{equation}
which in turn imply the kinematic relation
\begin{eqnarray}
\label{kcdotqrel}
2 k \cdot q = M_Z^2\,.
\end{eqnarray}
The polarization vectors for the photon and $Z$ satisfy
\begin{eqnarray}
\varepsilon^\mu_Z (p) p_\mu &=& 0 \,,
\label{e.p=0} \\ 
\varepsilon^\nu (k) k_\nu &=& 0 \,.
\label{e.k=0} 
\end{eqnarray}
The polarization tensor for the graviton satisfies the analogous
relations,
\begin{eqnarray}
{\cal E}^{\lambda\rho}(q) q_\lambda = 0 \,, \qquad
{\cal E}^{\lambda\rho} (q) q_\rho = 0 \,,
\label{e.q=0}
\end{eqnarray}
and in addition it is symmetric and traceless, i.e.,
\begin{eqnarray}
{\cal E}^{\lambda\rho} &=& {\cal E}^{\rho\lambda} \,,
\label{symmeps} \\ 
{\cal E}^{\lambda\rho} \eta_{\lambda\rho} &=& 0 \,.
\label{traceeps}
\end{eqnarray}

\Eq{symmeps} implies that $\T(1/)$ can be defined such that
\begin{equation}
\label{symmT}
\T(1/) = \T(1/N)_{\rho\lambda\mu\nu} \,.
\end{equation}
In addition, the conservation of the electromagnetic current and the
energy-momentum tensor, which are consequences of the electromagnetic
and gravitational gauge invariance, imply some additional properties
of $\T(1/)$ which will be useful in the explicit calculation of the
amplitude.  In the rest of this section, we explore the consequences
of these conditions.

The fact that the $Z$ and the graviton are electrically neutral
has two implications\footnote{When there are charged
particles, \Eqs{kT=0}{Ttaylor} are not valid. A relation analogous
to \Eq{kT=0} holds when the charged particles are on-shell,
but in that case the tree-level contributions
(Born diagrams) render the amplitude singular at $k = 0$.}:
(i) the conservation of the electromagnetic current yields the condition
\begin{eqnarray}
\label{kT=0}
k^\nu \T(1/) = 0 \,,
\end{eqnarray}
(ii) the absence of tree-level diagrams implies that $\T(1/)$ can be
expanded around $k = 0$.  We exploit these properties by writing
\begin{eqnarray}
\label{Ttaylor}
\T(1/) = {\cal T}^0_{\lambda\rho\mu\nu} + 
k^\alpha {\cal T}^1_{\lambda\rho\mu\nu\alpha} \,,
\end{eqnarray}
where ${\cal T}^0_{\lambda\rho\mu\nu}$ is independent of $k$.  Since
\Eq{kT=0} must be satisfied for all $k$, it implies that
\begin{eqnarray}
k^\nu {\cal T}^0_{\lambda\rho\mu\nu} & = & 0 \,,\\
k^\nu k^\alpha {\cal T}^1_{\lambda\rho\mu\nu\alpha} & = & 0 \,,
\end{eqnarray}
which in turn imply
\begin{eqnarray}
{\cal T}^0_{\lambda\rho\mu\nu} & = & 0\,, \nonumber\\
{\cal T}^1_{\lambda\rho\mu\nu\alpha} & = & \mbox{antisymmetric in
  $\nu\leftrightarrow\alpha$}\,.
\end{eqnarray}
Therefore, the amplitude can be written as
\begin{eqnarray}
\label{emform}
\mathscr M = {\cal E}^{\lambda\rho*} \varepsilon^\mu_Z f^{\nu\alpha\ast} 
t_{\lambda\rho\mu\nu\alpha}(q,k)\,,
\end{eqnarray}
where we have defined
\begin{eqnarray}
f_{\nu\alpha} \equiv k_\nu \varepsilon_\alpha - k_\alpha
\varepsilon_\nu \,. 
\end{eqnarray}
and $t_{\lambda\rho\mu\nu\alpha}$ is some undetermined tensor.

In analogous fashion, the conservation of the energy-momentum tensor
yields the condition
\begin{eqnarray}
q^\lambda j_{\lambda\rho} & = & 0\,,\nonumber\\
q^\rho j_{\lambda\rho} & = & 0 \,,
\label{q.j=0}
\end{eqnarray}
where we have defined
\begin{eqnarray}
j_{\lambda\rho} = \varepsilon^{\nu\ast} \varepsilon_Z^\mu
\T(1/) (q,k)\,.
\label{j}
\end{eqnarray}
Mimicking the argument of the electromagnetic case, we write
\begin{eqnarray}
\label{jdef}
j_{\lambda\rho} = j^0_{\lambda\rho} + 
j^1_{\lambda\rho\sigma}q^\sigma + 
j^2_{\lambda\rho\sigma\tau}q^\sigma q^\tau \,,
\end{eqnarray}
where $j^0_{\lambda\rho}$ and $j^1_{\lambda\rho\sigma}$ are
independent of $q$.  As before, it follows that the
transversality condition requires
\begin{eqnarray}
\label{cond1}
j^0_{\lambda\rho} & = & 0\,, \nonumber\\
j^1_{\lambda\rho\sigma} & = & -j^1_{\sigma\rho\lambda} \,.
\end{eqnarray}
In addition, \Eq{symmT} implies that
\begin{equation}
\label{cond2}
j^1_{\rho\lambda\sigma} = j^1_{\lambda\rho\sigma} \,.
\end{equation}
By applying Eqs.\ (\ref{cond1}) and (\ref{cond2}) successively and
repeatedly, we arrive at
\begin{equation}
j^1_{\lambda\rho\sigma} = -j^1_{\lambda\rho\sigma} \,,
\end{equation}
or in other words $j^1_{\lambda\rho\sigma} = 0$.

Considering now the structure of $j^2_{\lambda\rho\sigma\tau}$, we can see
that it has the following symmetry properties:
\begin{enumerate}\itemsep=0pt

\item\label{c.lr} It must be symmetric in the indices $\lambda,\rho$
  because of \Eq{symmT}.

\item\label{c.st} It must be symmetric in the indices $\sigma,\tau$
  because of its definition in \Eq{jdef}.

\item\label{c.lrst} Because of \Eq{q.j=0}, it must be antisymmetric when
we interchange anyone of the indices ($\lambda,\rho$) with anyone of
the indices ($\tau,\sigma$).

\end{enumerate}
Moreover, any term in $j^2_{\lambda\rho\sigma\tau}$ containing
$q_\lambda$ or $q_\rho$ vanishes in the amplitude because of the
on-shell transversality conditions of \Eq{e.q=0}, and any term
containing $q_\sigma$ or $q_\tau$ vanishes as well in \Eq{jdef} since
$q^2=0$.  Thus none of the Lorentz indices $\lambda,\rho,\tau,\sigma$
can be attached to the momentum $q$. Remembering \Eq{emform}, we know
that $j^2_{\lambda\rho\tau\sigma}$ must have a factor of
$f_{\alpha\beta}$, and of course it also involves a factor of
$\varepsilon_Z^\mu$. Thus, combining these pieces of information, and
excluding for the moment possible terms involving the Levi-Civita
tensor, it follows that $j^2_{\lambda\rho\tau\sigma}$ must be of the
form
\begin{eqnarray}
j^2_{\lambda\rho\tau\sigma} &=& F
f_{\lambda\tau}^\ast Z_{\rho\sigma} + (\lambda \leftrightarrow \rho)\,, 
\label{qqfZ}
\end{eqnarray}
where $F$ is some undetermined scalar, and we have introduced the notation
\begin{eqnarray}
Z_{\rho\sigma} = k_\rho \varepsilon^Z_\sigma - k_\sigma \varepsilon^Z_\rho\,, 
\end{eqnarray}
or equivalently
\begin{eqnarray}
Z_{\rho\sigma} = p_\rho \varepsilon^Z_\sigma - p_\sigma \varepsilon^Z_\rho\,.
\end{eqnarray}
The terms involving the Levi-Civita tensor are most easily enumerated
by noticing that all the requirements stated above for
$j^2_{\lambda\rho\tau\sigma}$ are satisfied if $f_{\alpha\beta}$
or $Z_{\alpha\beta}$ in \Eq{qqfZ} is replaced by its dual
\begin{eqnarray}
\widetilde f_{\alpha\beta} & = & \frac12
\epsilon_{\alpha\beta\mu\nu}f^{\mu\nu} 
\,,\nonumber\\
\widetilde Z_{\alpha\beta} & = & \frac12
\epsilon_{\alpha\beta\mu\nu}Z^{\mu\nu} \,,
\end{eqnarray}
respectively. Thus the most general form for the amplitude is
\begin{eqnarray}
\mathscr M = {\cal E}^{\lambda\rho\ast} q^\sigma q^\tau 
\Big[ F f_{\lambda\tau}^\ast Z_{\rho\sigma} + 
F_1 \widetilde f_{\lambda\tau}^\ast  Z_{\rho\sigma} + 
F_2 f_{\lambda\tau}^\ast \widetilde Z_{\rho\sigma} \Big] 
+ (\lambda \leftrightarrow \rho) \,,
\label{generalform}
\end{eqnarray}
where $F_1$ and $F_2$ are two additional Lorentz scalars.  Notice also
that a term involving both $\tilde f$ and $\tilde Z$ is not included,
since the product of two epsilon tensors can be written without it,
and all such terms are already exhausted in \Eq{generalform}.

It is convenient to write the expression for $\T(1/)$ that follows
from the form of the amplitude given in \Eq{generalform}. Using
the definition of \Eq{M} we get
\begin{eqnarray}
\T(1/) &=& 
\Big( F (k_\lambda q_\nu - k \cdot q \eta_{\nu\lambda}) 
   (k_\rho q_\mu - k\cdot q \eta_{\mu\rho}) 
\nonumber\\*  && 
+ F_1 [qk]_{\lambda\nu} (k_\rho q_\mu - k \cdot q \eta_{\mu\rho}) + 
F_2 (k_\lambda q_\nu - k \cdot q \eta_{\nu\lambda})
[qk]_{\rho\mu} \Big) 
+ (\lambda \leftrightarrow \rho) \,,
\label{Tgeneral}
\end{eqnarray}
where we have used the shorthand notation
\begin{eqnarray}
[qk]_{\mu\nu} \equiv \varepsilon_{\mu\nu\alpha\beta} q^\alpha k^\beta \,.
\end{eqnarray}
%

\section{Diagrams with fermion loops}\label{s:fl}
\subsection{Diagrams and Feynman rules}
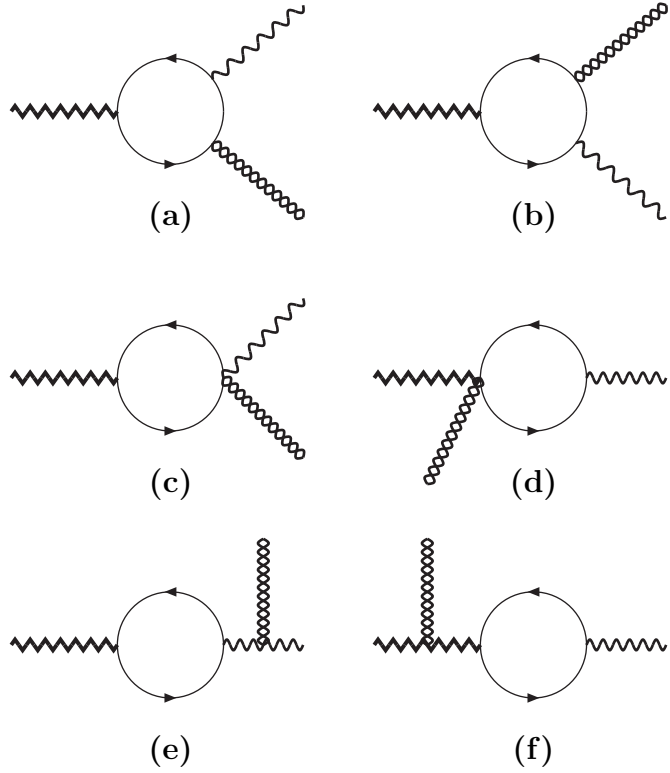
\begin{figure}[t]
\newcounter{subd}
\renewcommand{\thesubd}{\alph{subd}}
\def\loopdiag#1!!{\stepcounter{subd}
\begin{picture}(130,100)(-70,-50)
\ArrowArc(0,0)(20,0,180) 
\ArrowArc(0,0)(20,180,360)
\SetWidth{1.5} 
\ZigZag(-60,0)(-20,0)27
\SetWidth{1}
#1
\Text(0,-40)[]{\large\bf(\thesubd)}
\end{picture}}
\begin{center}

\loopdiag
\Photon(16,12)(50,40)26
\Graviton(16,-12)(50,-40)26
!!
\loopdiag
\Graviton(16,12)(50,40)26
\Photon(16,-12)(50,-40)26
!!

\loopdiag
\Graviton(20,0)(50,-30)26
\Photon(20,0)(50,30)26
!!
\loopdiag
\Graviton(-20,0)(-40,-40)26
\Photon(20,0)(50,0)26
!!

\loopdiag
\Graviton(35,0)(35,40)26
\Photon(20,0)(50,0)26
!!
\loopdiag
\Graviton(-40,0)(-40,40)26
\Photon(20,0)(50,0)26
!!
\end{center}
\caption{1-loop diagrams for the process $Z\to\gamma+\grav$ involving
  fermions in the loop.  In the external lines, the thick saw-tooth
  lines refer to the $Z$-boson, the thin wavy lines to the photon, and
  the braided lines to the graviton.}\label{f:fdiags}
\end{figure}

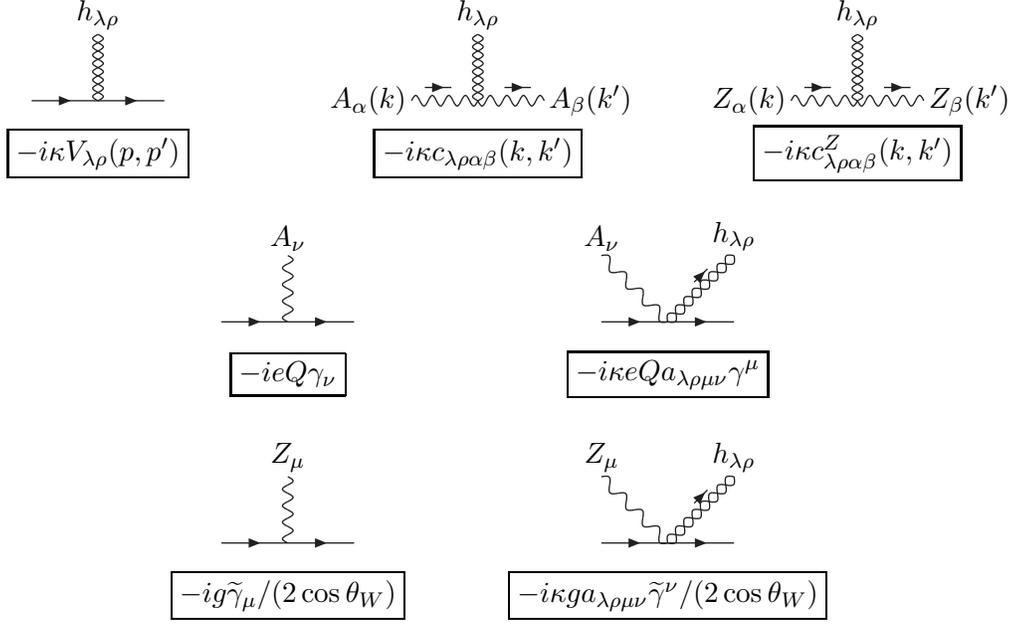
\begin{figure}[t]
\begin{center}
\begin{picture}(140,40)(-70,-25)
\ArrowLine(-25,0)(0,0)
\ArrowLine(0,0)(25,0)
\Graviton(0,25)(0,0)24
\Text(0,27)[b]{$h_{\lambda\rho}$}
\Text(0,-20)[]{\framebox{$-i\kappa V_{\lambda\rho}(p,p')$}}
\end{picture}
\begin{picture}(140,40)(-70,-25)
\Photon(-25,0)(0,0)24
\ArrowLine(-20,5)(-10,5)
\Text(-27,0)[r]{$A_\alpha(k)$}
\Photon(25,0)(0,0){-2}4
\ArrowLine(10,5)(20,5)
\Text(27,0)[l]{$A_\beta(k')$}
\Graviton(0,25)(0,0)24
\Text(0,27)[b]{$h_{\lambda\rho}$}
\Text(0,-20)[]{\framebox{$-i\kappa c_{\lambda\rho\alpha\beta}(k,k')$}} 
\end{picture}
\begin{picture}(140,40)(-70,-25)
\Photon(-25,0)(0,0)24
\ArrowLine(-20,5)(-10,5)
\Text(-27,0)[r]{$Z_\alpha(k)$}
\Photon(25,0)(0,0){-2}4
\ArrowLine(10,5)(20,5)
\Text(27,0)[l]{$Z_\beta(k')$}
\Graviton(0,25)(0,0)24
\Text(0,27)[b]{$h_{\lambda\rho}$}
\Text(0,-20)[]{\framebox{$-i\kappa c^Z_{\lambda\rho\alpha\beta}(k,k')$}} 
\end{picture}
\\[15mm] 
\begin{picture}(140,40)(-70,-25)
\ArrowLine(-25,0)(0,0)
\ArrowLine(0,0)(25,0)
\Photon(0,25)(0,0)24
\Text(0,27)[b]{$A_\nu$}
\Text(0,-20)[]{\framebox{$-ieQ \gamma_\nu$}}
\end{picture}
\begin{picture}(140,40)(-70,-25)
\ArrowLine(-25,0)(0,0)
\ArrowLine(0,0)(25,0)
\Photon(-25,25)(0,0)24
\Text(-25,27)[b]{$A_\nu$}
\Graviton(25,25)(0,0)24
\Text(25,27)[b]{$h_{\lambda\rho}$}
\ArrowLine(10,15)(15,20)
\Text(0,-20)[]{\framebox{$-i\kappa eQ a_{\lambda\rho\mu\nu} \gamma^\mu$}} 
\end{picture}
\\[15mm] 
\begin{picture}(140,40)(-70,-25)
\ArrowLine(-25,0)(0,0)
\ArrowLine(0,0)(25,0)
\Photon(0,25)(0,0)24
\Text(0,27)[b]{$Z_\mu$}
\Text(0,-20)[]{\framebox{$-ig \gtil_\mu/(2\cos\theta_W)$}}
\end{picture}
\begin{picture}(140,40)(-70,-25)
\ArrowLine(-25,0)(0,0)
\ArrowLine(0,0)(25,0)
\Photon(-25,25)(0,0)24
\Text(-25,27)[b]{$Z_\mu$}
\Graviton(25,25)(0,0)24
\Text(25,27)[b]{$h_{\lambda\rho}$}
\ArrowLine(10,15)(15,20)
\Text(0,-20)[]{\framebox{$-i\kappa g a_{\lambda\rho\mu\nu}
    \gtil^\nu/(2\cos\theta_W)$}} 
\end{picture}
\end{center}

\caption{Notations for Feynman rules for couplings that appear in
  \fig{f:fdiags}.  The charge of the fermion is $eQ_f$.  Various
  symbols appearing in this figure have been explained in the
  text.}\label{f:frules}
\end{figure}

As already mentioned in the Introduction, we consider first the
fermion loop diagrams, which are shown in \fig{f:fdiags}. The Feynman
rules for the various vertices that appear there have been given in
the literature \cite{Choi:1994ax, Shim:1995ap, Nieves:1998xz,
Nieves:1999rt, Nieves:2000dc}, and for convenience they are summarized
in \fig{f:frules}.
Any fermion $f$ that circulates in the loop must be electrically
charged since it is attached to the photon line, and we denote its
charge by $e Q_f$ where $e$ is the charge of the positron. The
gravitational vertex function of the the fermion is given by
\begin{eqnarray}
V_{\lambda\rho} (p,p') = \frac14 \Big[\gamma_\lambda (p+p')_\rho +
  \gamma_\rho (p+p')_\lambda \Big] - \frac12 \eta_{\lambda\rho}
  \Big[\slash p + \slash p' - 2m_f \Big] \,,
\label{V}
\end{eqnarray}
which can also be written as
\begin{eqnarray}
V_{\lambda\rho} (p,p') = -\frac12 a_{\lambda\rho\alpha\beta}
(p+p')^\alpha \gamma^\beta + m_f \eta_{\lambda\rho} \,,
\label{Va}
\end{eqnarray}
where
\begin{eqnarray}
a_{\lambda\rho\mu\nu} = \eta_{\lambda\rho} \eta_{\mu\nu} - \frac12
\eta_{\lambda\rho\mid\mu\nu} \,.
\label{a}
\end{eqnarray}
In writing this form, we have used the shorthand notation
\begin{eqnarray}
\eta_{\lambda\rho\mid\mu\nu} \equiv \eta_{\lambda\mu} \eta_{\rho\nu} +
\eta_{\lambda\nu} \eta_{\rho\mu} \,.
\label{eta4}
\end{eqnarray}
The tensor defined in \Eq{a} also appears in the coupling of a fermion
bilinear to a gauge boson and the graviton.  For example, if we denote
Feynman rule for the $Z$-boson coupling to a fermion by
$-ig\gtil_\mu/(2\cos\theta_W)$, where
\begin{eqnarray}
\gtil_\mu \equiv \gamma_\mu (X_f + Y_f \gamma_5) \,,
\label{gtil}
\end{eqnarray}
the fermion-$Z$-$\grav$ vertex is given by $-i\kappa g
a_{\lambda\rho\nu\nu} \gtil^\nu/(2\cos\theta_W)$, as indicated in
\fig{f:frules}.  In the standard model,
\begin{eqnarray}
X_f &=& T_{Lf} - 2Q_f\sin^2\theta_W  \,, \nonumber\\
Y_f &=& - T_{Lf} \,,
\end{eqnarray}
where $T_{Lf}$ is the eigenvalue of the diagonal generator of SU(2)
acting on the left-chiral component of the fermion.

As emphasized earlier, we employ the unitary gauge for the gauge bosons,
which means that their propagators and the gravitational couplings
are determined using the canonical form of the kinetic term in the
bilinear part of the Lagrangian. The gravitational vertex function
$c_{\lambda\rho\alpha\beta}$ of the photon is then given by
\begin{eqnarray}
c_{\lambda\rho\alpha\beta} (k,k') &=& \eta_{\lambda\rho} (
\eta_{\alpha\beta} k 
\cdot k' - k'_\alpha k_\beta ) - \eta_{\alpha\beta} (k_\lambda k'_\rho +
k'_\lambda k_\rho ) \nonumber\\*
&& + k_\beta (\eta_{\lambda\alpha} k'_\rho +
\eta_{\rho\alpha} k'_\lambda)
+ k'_\alpha (\eta_{\lambda\beta} k_\rho +
\eta_{\rho\beta} k_\lambda) \nonumber\\*
&& - k \cdot k' (\eta_{\lambda\alpha} \eta_{\rho\beta} +
\eta_{\lambda\beta} \eta_{\rho\alpha}) \,,
\label{c(old)}
\end{eqnarray}
and it is is useful to note that it satisfies
\begin{eqnarray}
k^\alpha c_{\lambda\rho\alpha\beta}(k,k') = k'^\beta 
c_{\lambda\rho\alpha\beta}(k,k^\prime) = 0 \,.
\label{transvC}
\end{eqnarray}
For our present purposes it is convenient to write it in the more compact form
\begin{eqnarray}
c_{\lambda\rho\alpha\beta} (k,k') &=& \eta_{\lambda\rho} (
\eta_{\alpha\beta} k 
\cdot k' - k'_\alpha k_\beta^{\phantom '} ) - \Big(
\eta_{\lambda\rho\mid\alpha\beta,\mu\nu} -
\eta_{\lambda\rho\mid\alpha\nu,\beta\mu} \Big) k^\mu k'^\nu \,,
\label{c}
\end{eqnarray}
where we have introduced the notation
\begin{eqnarray}
\eta_{\lambda\rho\mid\alpha\beta,\mu\nu} \equiv
\eta_{\lambda\rho\mid\alpha\beta} \eta_{\mu\nu} +
\eta_{\lambda\rho\mid\mu\nu} \eta_{\alpha\beta} \,,
\label{eta6}
\end{eqnarray}
with $\eta_{\lambda\rho\mid\mu\nu}$ defined in \Eq{eta4}.
Similarly, the gravitational vertex function of the $Z$-boson is given by
\begin{eqnarray}
c^Z_{\lambda\rho\alpha\beta} (k,k') = c_{\lambda\rho\alpha\beta}
(k,k') - M_Z^2 a'_{\lambda\rho\alpha\beta} \,,
\label{cZ}
\end{eqnarray}
where
\begin{eqnarray}
a'_{\lambda\rho\mu\nu} = \eta_{\lambda\rho} \eta_{\mu\nu} - 
\eta_{\lambda\rho\mid\mu\nu} \,.
\label{a'}
\end{eqnarray}
With this choice of vertices, the $Z$-propagator to be used is then
given by
\begin{eqnarray}
D^{\alpha\beta}_Z (k) = {1 \over k^2-M_Z^2} \left(
-\eta^{\alpha\beta} + {k^\alpha k^\beta \over M_Z^2} \right) \,.
\label{DZ}
\end{eqnarray}

For the photon propagator $D^{\alpha\beta}(k)$, the equation of motion
that follows from the Lagrangian given in \Eq{L0photon} determines
only the transverse part of the solution, leaving an undetermined
longitudinal part. That is, in the presence of a source $J_\mu$, the
vector potential is given by
\begin{equation}
\label{eqmotionsol}
A_\alpha = D_{\alpha\beta} J^\beta\,,
\end{equation}
where
\begin{equation}
D^{\alpha\beta}(k) = \frac{1}{k^2}\left(
-\eta^{\alpha\beta} + {k^\alpha k^\beta \over k^2} \right) + 
D_L k^\alpha k^\beta\,,
\end{equation}
with $D_L$ being an undetermined scalar function of $k$.
The consistency of the equation of motion requires that $J_\alpha$ be
conserved, which means that $D_L$ drops out in \Eq{eqmotionsol},
and in fact that we can set
\begin{equation}
D^{\alpha\beta} (k) = \frac{-\eta^{\alpha\beta}}{k^2}
\label{DA}
\end{equation}
in the solution.  The counterpart to this result in the context of our
calculation is that, since the photon propagator enters only in the
diagram in which the graviton is attached to the external photon line,
then by virtue \Eq{transvC} for practical purposes the propagator can
be taken as given in \Eq{DA}, which is what we adopt.

\subsection{The amplitude}
We denote by $\T(f/1)$ the contribution to $\T(1/)$ from one fermion
$f$ in the loop. Then writing it in the form
\begin{eqnarray}
\T(f/1) = {\kappa eg \over 2\cos\theta_W} Q_f \T(f/0) \,,
\label{Tferm}
\end{eqnarray}
the contributions from the various diagrams to $\T(f/0)$ are given by
\begin{eqnarray}
\T(f/a) &=& i\lint l \Tr \Big[ \gtil_\mu S(l-k) \gamma_\nu S(l)
  V_{\lambda\rho} (l+q,l) S(l+q) \Big] \,, \nonumber\\ 
\T(f/b) &=& i\lint l \Tr \Big[ \gtil_\mu S(l) V_{\lambda\rho} (l+q,l)
  S(l+q) \gamma_\nu S(l+p) \Big] \,, \nonumber\\ 
\T(f/c) &=& 
a_{\lambda\rho\alpha\nu} {\Pi_\mu}^\alpha (p) \,, \nonumber\\ 
\T(f/d) &=& 
a_{\lambda\rho\mu\alpha} {\Pi^{\alpha}}_\nu(k) \,, \nonumber\\  
\T(f/e) &=& 
c_{\lambda\rho\alpha\nu} (p,k) D^{\alpha\beta} (p)
  \Pi_{\mu\beta}(p) \,,  \nonumber\\ 
\T(f/f) &=& 
c^Z_{\lambda\rho\mu\alpha} (p,k) D^{\alpha\beta}_Z (k)
  \Pi_{\beta\nu}(k) \,,
\label{T(a..f)}
\end{eqnarray}
where
\begin{eqnarray}
\Pi_{\mu\nu} (k) = i \lint l \Tr \Big[ \gtil_\mu S(l)
  \gamma_\nu S(l+k) \Big] \,.
\label{PigamZ}
\end{eqnarray}
$\Pi_{\mu\nu} (k)$, which is immediately recognized to be the fermion
contribution to the $\gamma Z$ polarization mixing tensor, satisfies
the transversality condition
\begin{eqnarray}
k^\nu\Pi_{\mu\nu}(k) = 0 \,,
\label{transvpi}
\end{eqnarray}
which is a consequence of electromagnetic gauge invariance
and which can be proved explicitly by means of the elementary identity
\begin{eqnarray}
k^\nu S(l) \gamma_\nu S(l+k) = S(l) - S(l+k) \,.
\label{Sk/S}
\end{eqnarray}
This property of $\Pi_{\mu\nu}(k)$ implies that it is of the form 
\begin{eqnarray}
\Pi_{\mu\nu}(k) = \left(\eta_{\mu\nu} - \frac{k_\mu
  k_\nu}{k^2}\right)\Pi(k)\,, 
\label{Piform}
\end{eqnarray}
where $\Pi(k)$ is given by a (logarithmically divergent) integral
that can be obtained from \Eq{PigamZ} but whose precise value, as we 
will see, is not relevant for our calculation.

\subsection{The electromagnetic transversality condition}
\Eq{transvpi} implies
\begin{eqnarray}
\label{kTdf}
k^\nu \T(f/d) =  k^\nu \T(f/f) =  0 \,,
\end{eqnarray}
and using \Eq{transvC}
\begin{eqnarray}
\label{kTe}
k^\nu \T(f/e) = 0 \,.
\end{eqnarray}
Then using \Eq{Sk/S} and remembering the momentum conservation equation
\Eq{p=k+q}, we obtain
\begin{eqnarray}
k^\nu \T(f/a) &=& i\lint l \Tr \Big[ \gtil_\mu \Big\{ S(l-k) - S(l) \Big\}
  V_{\lambda\rho} (l,l+q) S(l+q) \Big] \,, \nonumber\\ 
k^\nu \T(f/b) &=& i\lint l \Tr \Big[ \gtil_\mu S(l) V_{\lambda\rho}
  (l,l+q) \Big\{ S(l+q) - S(l+p) \Big\} \Big] \,,
\end{eqnarray}
and therefore
\begin{eqnarray}
k^\nu \T(f/a+b) = i\lint l \Tr \Big[ \gtil_\mu S(l-k) 
  V_{\lambda\rho} (l+q,l) S(l+q) 
- \gtil_\mu S(l) V_{\lambda\rho} (l+q,l)
   S(l+p) \Big] \,.
\end{eqnarray}
Changing the dummy loop momentum in the first term this can be written as
\begin{eqnarray}
k^\nu \T(f/a+b) = i\lint l \Tr \Big[ \gtil_\mu S(l) 
  \Big\{ V_{\lambda\rho} (l+p,l+k) - V_{\lambda\rho} (l+q,l) \Big\}
   S(l+p) \Big] \,,
\end{eqnarray}
and using the identity given in \Eq{V-V}
\begin{eqnarray}
k^\nu \T(f/a+b) &=& \null - a_{\lambda\rho\alpha\nu} k^\nu
   {\Pi_\mu}^\alpha (p) \nonumber\\*
&=& \null - k^\nu \T(f/c)\,. 
\label{kTabc}
\end{eqnarray}
Using \Eqss{kTdf}{kTe}{kTabc}, it follows that
\begin{equation}
k^\nu \T(f/0) = 0 \,,
\end{equation}
which establishes the electromagnetic gauge invariance of this set of
diagrams, in the sense that their contribution to the amplitude
satisfies the requirement due to the conservation of the
electromagnetic current.

\subsection{The gravitational transversality condition}
In similar fashion, we now establish that
\begin{equation}
\label{q.Tf=0}
q^\lambda \varepsilon^{\nu\ast}(k)\varepsilon^\mu_Z (p) \T(f/1) = 0\,.
\end{equation}
In contrast with the electromagnetic case, here both the $Z$ and the
photon are assumed to be on-shell.

Using the identity
\begin{eqnarray}
S(l) q^\lambda V_{\lambda\rho}(l+q,l) S(l+q) 
&=& l_\rho S(l) - (l_\rho + q_\rho) S(l+q) \nonumber\\* 
&& + \frac18 S(l) \Big( \gamma_\rho \slash q - \slash q \gamma_\rho
\Big) - \frac18 \Big( \gamma_\rho \slash q - \slash q 
\gamma_\rho \Big) S(l+q) \,,
\label{SqVS}
\end{eqnarray}
together with the Dirac-matrix identity
\begin{eqnarray}
\gamma_\alpha \gamma_\beta \gamma_\rho + \gamma_\rho \gamma_\beta
\gamma_\alpha = 2 \Big( \eta_{\alpha\beta} \gamma_\rho +
\eta_{\beta\rho} \gamma_\alpha - \eta_{\alpha\rho} \gamma_\beta \Big)
\,, 
\label{g+g+g}
\end{eqnarray}
and making judicious shifts in the loop momenta in various terms we obtain
\begin{eqnarray}
q^\lambda \T(f/a+b) &=&  k_\rho \Pi_{\mu\nu} (k) -
  p_\rho \Pi_{\mu\nu} (p) 
+ \frac12 \Big( q_\nu \Pi_{\mu\rho} (p) - \eta_{\nu\rho} q^\alpha
  \Pi_{\mu\alpha} (p) \Big) 
  \nonumber\\* 
&& + \frac12 \Big( q_\mu \Pi_{\rho\nu} (k) - \eta_{\mu\rho} q^\alpha
  \Pi_{\alpha\nu} (k) \Big) \,.
\end{eqnarray}
The definition of $a_{\lambda\rho\mu\nu}$ given in \Eq{a} allows us to
write
\begin{eqnarray}
q^\lambda \T(f/c) &=& q_\rho \Pi_{\mu\nu} (p)
- \frac12 q_\nu \Pi_{\mu\rho} (p) -
\frac12 \eta_{\nu\rho} q^\alpha \Pi_{\mu\alpha} (p) \,,\\ 
q^\lambda \T(f/d) &=& q_\rho \Pi_{\mu\nu} (k) 
- \frac12 q_\mu \Pi_{\rho\nu} (k) -
\frac12 \eta_{\mu\rho} q^\alpha \Pi_{\alpha\nu} (k) \,,
\end{eqnarray}
and therefore
\begin{eqnarray}
\T(f/a+b+c+d) &=& p_\rho \Pi_{\mu\nu} (k) -
  k_\rho \Pi_{\mu\nu} (p)  - \eta_{\nu\rho} q^\alpha
  \Pi_{\mu\alpha} (p) - \eta_{\mu\rho} q^\alpha
  \Pi_{\alpha\nu} (k) \,.
\label{q.T(abcd)}
\end{eqnarray}

For diagram \fig{f:fdiags}e, using \Eq{p=k+q} we first find that
\begin{eqnarray}
q^\lambda \varepsilon^{\nu\ast}(k)\varepsilon^{\mu}(p)
c_{\lambda\rho\alpha\nu} (p,k) =  \varepsilon^{\nu\ast}(k)
\varepsilon^{\mu}(p)\left[(k_\alpha
\eta_{\rho\nu} - k_\rho \eta_{\alpha\nu}) p^2 +  (q_\nu k_\rho -
\eta_{\rho\nu} k \cdot q) p_\alpha\right] \,,
\label{q.C}
\end{eqnarray}
where we have used \Eq{e.k=0} and the on-shell photon condition $k^2=0$.
By the transversality property of $\Pi_{\mu\nu}$ we can use
\begin{eqnarray}
p_\alpha D^{\alpha\beta}(p) \Pi_{\mu\beta} (p) 
\propto p_\alpha {\Pi^\alpha}_\mu (p)  = 0\,,
\end{eqnarray}
which implies that
\begin{eqnarray} 
q^\lambda \varepsilon^{\nu\ast}(k)\varepsilon^\mu_Z (p)
\T(f/e) &=& \varepsilon^{\nu\ast}(k)\varepsilon^\mu_Z (p)
\left[k_\rho \Pi_{\mu\nu} (p) 
- \eta_{\nu\rho} k^\alpha \Pi_{\mu\alpha} (p)\right] \,.
\label{q.Te}
\end{eqnarray}
Similarly, for the expression involving diagram of
\fig{f:fdiags}f, we find
\begin{eqnarray}
q^\lambda \varepsilon^{\nu\ast}(k)\varepsilon^\mu_Z(p)
c^Z_{\lambda\rho\mu\alpha} (p,k) = 
\varepsilon^{\nu\ast}(k)\varepsilon^\mu_Z(p)\left[(p_\rho
\eta_{\mu\alpha} - p_\alpha \eta_{\mu\rho}) (k^2 - M_Z^2) 
+ (q_\mu p_\rho - \eta_{\rho\mu} p \cdot q) k_\alpha\right] \,,\nonumber\\
\label{q.CZ}
\end{eqnarray}
and using the $Z$-propagator from \Eq{DZ}
\begin{eqnarray}
q^\lambda \varepsilon^{\nu\ast}(k)\varepsilon^\mu_Z(p)
c^Z_{\lambda\rho\mu\alpha} (p,k) D^{\alpha\beta}_Z (k) =
\varepsilon^{\nu\ast}(k)\varepsilon^\mu_Z(p)\left[
\eta_{\mu\rho} q^\beta - \eta_\mu^\beta p_\rho\right] \,,
\label{q.CZ.D}
\end{eqnarray}
which gives
\begin{eqnarray}
q^\lambda \varepsilon^{\nu\ast}(k)\varepsilon^\mu_Z(p) \T(f/f) &=& 
\varepsilon^{\nu\ast}(k)\varepsilon^\mu_Z(p)\left[
\eta_{\mu\rho} q^\alpha \Pi_{\alpha\nu} (k) 
- p_\rho \Pi_{\mu\nu} (k)\right] \,.
\label{q.Tf}
\end{eqnarray}
Adding Eqs.\ (\ref{q.T(abcd)}), (\ref{q.Te}) and (\ref{q.Tf}), and
using \Eqs{p=k+q}{transvpi}, \Eq{q.Tf=0} is established.

\subsection{Calculation of the form factors}
\label{subsec:fermionscalc}
The fact that the contribution of the fermion loop diagrams to the
amplitude satisfies the electromagnetic and gravitational
transversality conditions implies that it must have the structure
given in \Eq{Tgeneral}.  In order to extract the corresponding
contribution to the form factors we need not evaluate in full the
integral expressions for the various diagrams, and instead we can
proceed as follows. The form factors $F_{1,2}$ are easily identified
as the coefficients of the terms containing the Levi-Civita tensor. As
we will see, no such terms appear so that these form factors are
zero. For the form factor $F$, we can fix our attention on the
contributions to just one of the terms that appear in
\Eq{Tgeneral}. To be specific, we choose $F k_\lambda k_\rho q_\mu
q_\nu$. Since, by \Eq{e.p=0},
\begin{eqnarray}
\varepsilon^\mu_Z (p) k_\mu = \null - \varepsilon^\mu_Z (p) q_\mu 
\label{k=-q}
\end{eqnarray}
such terms will arise from the terms in the integrals that contain a
factor of either $k_\lambda k_\rho q_\mu q_\nu$ or $k_\lambda k_\rho
k_\mu q_\nu$.  A term of either form will be called a $kkqq$ term for
the sake of brevity, and these are the only ones that we need to track
in the evaluation of the integrals.

We now consider the contribution from each diagram to that kind of
term, using the expressions given in \Eq{T(a..f)}. Since $\T(f/c)$ and
$\T(f/d)$ depend only on $p$ and $k$, respectively, neither one
contains a $kkqq$ term. Using \Eqs{DZ}{Piform} it is immediately
realized that $\T(f/e)$ and $\T(f/f)$ are proportional to
$c_{\lambda\rho\mu\nu}(p,k)$ and $c^Z_{\lambda\rho\mu\nu}(p,k)$,
respectively, neither one of which contains a $kkqq$ term as can be
seen simply by looking at their definitions given in \Eqs{c}{cZ}.

Thus, we are left with $\T(f/a)$ and $\T(f/b)$ which
using the graviton vertex in the form given in \Eq{Va} and remembering
\Eqs{symmeps}{traceeps}, can be written as
\begin{eqnarray}
\T(f/a) &=& i\lint l \; {l_\rho \Tr \Big[ \gtil_\mu (\slash l -
    \slash k + m_f) \gamma_\nu (\slash l + m_f) \gamma_\lambda (\slash
    l + \slash q + m_f) \Big] \over
    [(l-k)^2-m_f^2][(l+q)^2-m_f^2](l^2-m_f^2)} \,, 
    \nonumber\\ 
\T(f/b) &=& i\lint l \; {l_\rho \Tr \Big[ \gtil_\mu (\slash l - \slash
    q + m_f) \gamma_\lambda (\slash l + m_f) \gamma_\nu (\slash l +
    \slash k + m_f) \Big] \over
    [(l+k)^2-m_f^2][(l-q)^2-m_f^2](l^2-m_f^2)} \,.
\end{eqnarray}
Changing the integration variable from $l$ to $-l$ in the integral
for $\T(f/b)$, and then using the cyclic property of the trace,
together with the relations
\begin{eqnarray}
C^{-1}\gamma^\top_\mu C & = & -\gamma_\mu \nonumber\\
C^{-1}S^\top (\ell) C & = & S(-\ell) \,,
\end{eqnarray}
and similar ones, it follows that the $\gamma_\mu\gamma_5$
axial current coupling from $\T(f/a)$ and $\T(f/b)$ are opposite, while
the vector current term $\gamma_\mu$ is the same. Therefore,
\begin{eqnarray}
\T(f/a+b) &=& 2iX_f\lint l \; \frac{f_{\lambda\rho\mu\nu}(l)}
{[(l-k)^2-m_f^2][(l+q)^2-m_f^2](l^2-m_f^2)} \,,
\end{eqnarray}
where
\begin{equation}
f_{\lambda\rho\mu\nu}(l) = l_\rho \Tr \Big[ \gamma_\mu (\slash l -
    \slash k + m_f) \gamma_\nu (\slash l + m_f) \gamma_\lambda (\slash
    l + \slash q + m_f) \Big] \,.
\end{equation}
The absence of the $\gamma_5$ term implies that
there are no terms containing the Levi-Civita tensor,
so that there are no contributions to the form factors $F_{1,2}$.
Continuing to $D$ dimensions and parameterizing the integrals in the standard
fashion we obtain
\begin{eqnarray}
\T(f/a+b) = 4iX_f \lintd l \int_0^1 dx \int_0^{1-x} dy \;
  {f_a(l+xk-yq) \over \Big[ l^2 - m_f^2 + xy M_Z^2 \Big]^3} \,,
\end{eqnarray}
where we have used the on-shell relations given in
\Eqss{k^2=0}{q^2=0}{kcdotqrel}.  Evaluating the trace and focusing on
the $kkqq$ terms as described above, we get
\begin{eqnarray}
\T(f/a+b) = \null - 64iX_f q_\mu q_\nu k_\lambda k_\rho 
\lint l \int_0^1 dx
  \int_0^{1-x} dy \; 
  {x^2 y (1-x-y) \over \Big[ l^2 - m_f^2 + xy
  M_Z^2 \Big]^3} 
+ \cdots \,,
\label{f/a+b}
\end{eqnarray}
where we have continued back to four dimensions since 
the resulting integral is convergent and the ellipses indicate that
have omitted all the other terms.
Performing the momentum integration and recalling the
overall factors in \Eq{Tferm}, we finally find
the contribution from any fermion in the loop to
the form factor $F$ to be
\begin{eqnarray}
F^\di(f/0) = \null - {\kappa eg \over 2\pi^2\cos\theta_W} Q_f X_f I(m_f/M_Z)\,,
\label{Fferm}
\end{eqnarray}
where 
\begin{equation}
I(A) = \int_0^1 dx
  \int_0^{1-x} dy \; {x^2 y (1-x-y) \over A^2 - xy} \,.
\label{I}
\end{equation}

\begin{figure}
\setcounter{subd}{0}
\renewcommand{\thesubd}{\alph{subd}}
\def\loopdiag#1!!{\stepcounter{subd}
\begin{picture}(130,100)(-70,-60)
\PhotonArc(0,0)(20,0,360)2{20}
\SetWidth{1.5} 
\ZigZag(-60,0)(-20,0)27
\SetWidth{1}
#1
\Text(0,-40)[]{\large\bf(\thesubd)}
\end{picture}}
\begin{center}

\loopdiag
\Photon(16,12)(50,40)26
\Graviton(16,-12)(50,-40)26
!!
\loopdiag
\Graviton(16,12)(50,40)26
\Photon(16,-12)(50,-40)26
!!
\loopdiag
\Graviton(20,0)(50,-30)26
\Photon(20,0)(50,30)26
!!
\loopdiag
\Graviton(-20,0)(-40,-40)26
\Photon(20,0)(50,0)24
!!
\loopdiag
\Graviton(35,0)(35,40)26
\Photon(20,0)(50,0)26
!!
\loopdiag
\Graviton(-40,0)(-40,40)26
\Photon(20,0)(50,0)26
!!

\loopdiag
\Photon(-20,0)(-40,-40)26
\Graviton(20,0)(50,0)24
!!
\loopdiag
\Photon(-20,0)(-30,40)26
\Graviton(-20,0)(-30,-40)24
!!

\loopdiag
\Photon(-20,0)(-20,40)26
\Graviton(-20,30)(20,30)24
!!
\loopdiag
\Photon(-20,0)(-20,40)26
\Graviton(-40,0)(-40,-40)24
!!
\end{center}
\caption{1-loop diagrams for the process $Z\to\gamma+\grav$ involving
  charged gauge bosons in the loop.  In the external lines, the thick
  saw-tooth lines refer to the $Z$-boson, the thin wavy line to the
  photon, and the braided lines to the graviton.}\label{f:Wdiags}
\end{figure}
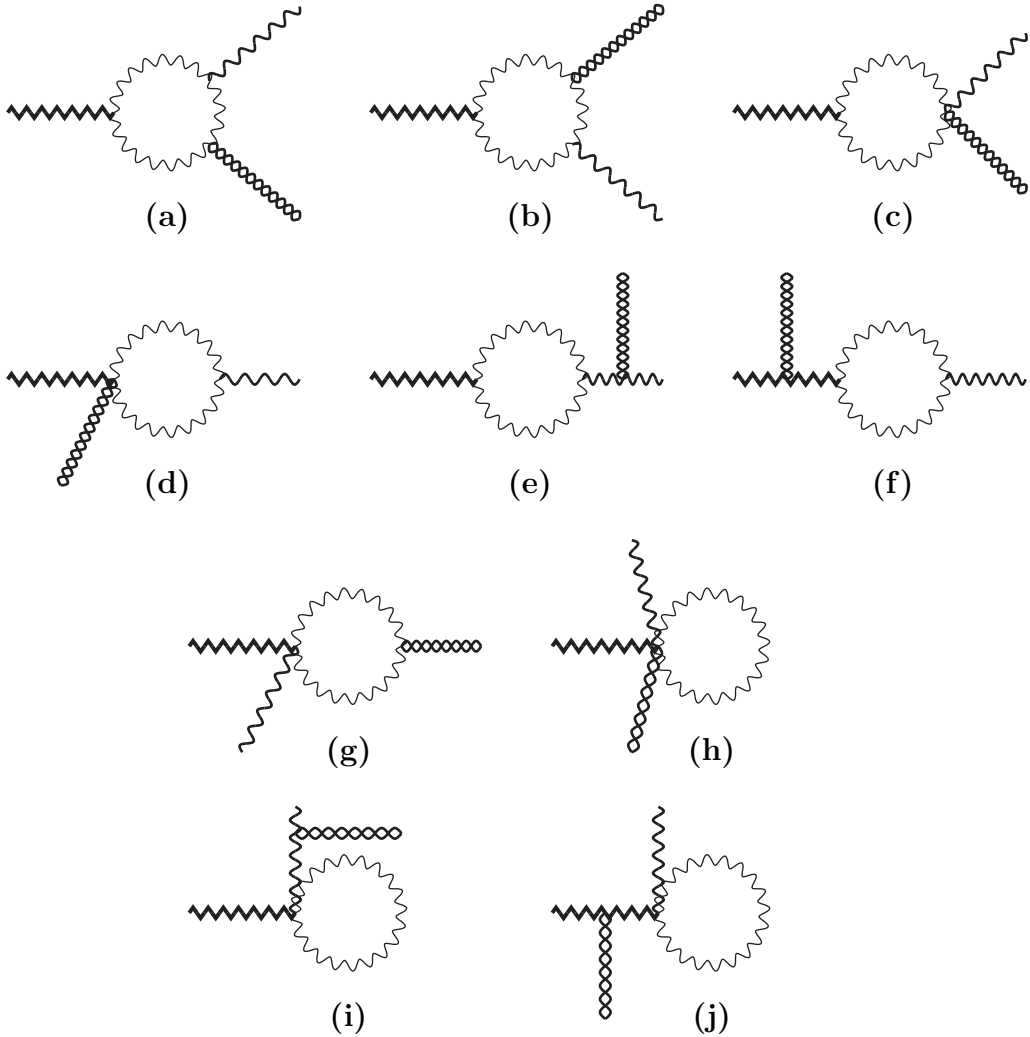
\section{Diagrams with $W$ loops}
\label{s:Wl}
\subsection{Diagrams and Feynman rules}
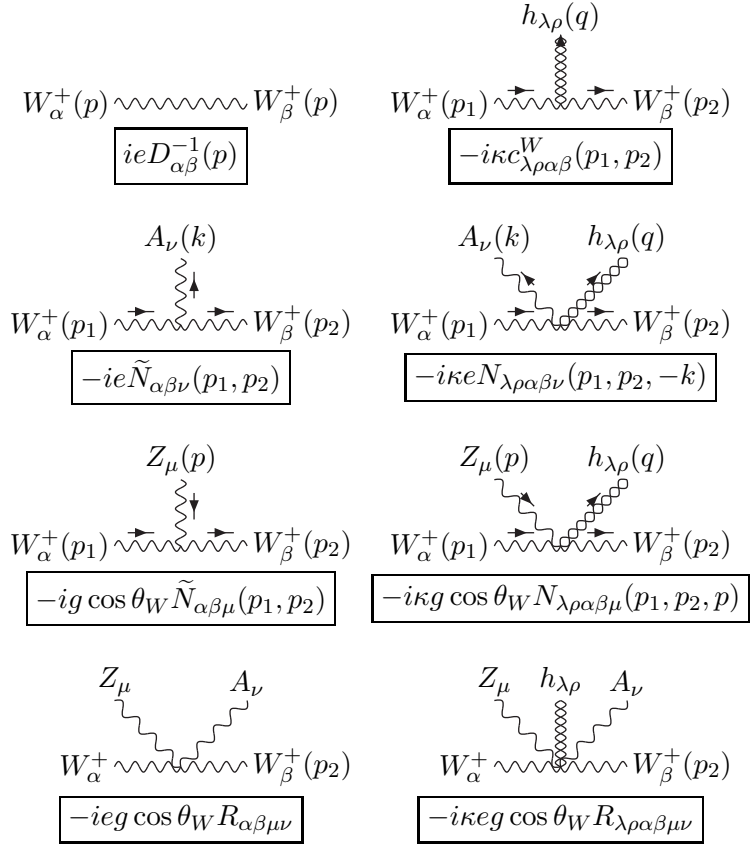
\begin{figure}[t]
\begin{center}
\begin{picture}(140,40)(-70,-25)
\Photon(-25,0)(25,0)28
\Text(-27,0)[r]{$W^+_\alpha(p)$}
\Text(27,0)[l]{$W^+_\beta(p)$}
\Text(0,-20)[]{\framebox{$ie D^{-1}_{\alpha\beta}(p)$}}
\end{picture}
\begin{picture}(140,40)(-70,-25)
\Photon(-25,0)(0,0)24
\ArrowLine(-20,5)(-10,5)
\Text(-27,0)[r]{$W^+_\alpha(p_1)$}
\Photon(25,0)(0,0){-2}4
\ArrowLine(10,5)(20,5)
\Text(27,0)[l]{$W^+_\beta(p_2)$}
\Graviton(0,25)(0,0)24
\Text(0,27)[b]{$h_{\lambda\rho}(q)$}
\ArrowLine(0,22)(0,26)
\Text(0,-20)[]{\framebox{$-i\kappa 
    c^W_{\lambda\rho\alpha\beta}(p_1,p_2)$}} 
\end{picture}
\\[15mm] 
\begin{picture}(140,40)(-70,-25)
\Photon(-25,0)(0,0)24
\ArrowLine(-20,5)(-10,5)
\Text(-27,0)[r]{$W^+_\alpha(p_1)$}
\Photon(25,0)(0,0){-2}4
\ArrowLine(10,5)(20,5)
\Text(27,0)[l]{$W^+_\beta(p_2)$}
\Photon(0,25)(0,0)24
\ArrowLine(5,10)(5,20)
\Text(0,27)[b]{$A_\nu(k)$}
\Text(0,-20)[]{\framebox{$-ie \Ntil_{\alpha\beta\nu}(p_1,p_2)$}}
\end{picture}
\begin{picture}(140,40)(-70,-25)
\Photon(-25,0)(0,0)24
\ArrowLine(-20,5)(-10,5)
\Text(-27,0)[r]{$W^+_\alpha(p_1)$}
\Photon(25,0)(0,0){-2}4
\ArrowLine(10,5)(20,5)
\Text(27,0)[l]{$W^+_\beta(p_2)$}
\Photon(-25,25)(0,0)24
\ArrowLine(-10,15)(-15,20)
\Text(-25,27)[b]{$A_\nu(k)$}
\Graviton(25,25)(0,0)24
\Text(25,27)[b]{$h_{\lambda\rho}(q)$}
\ArrowLine(10,15)(15,20)
\Text(0,-20)[]{\framebox{$-i\kappa e
     N_{\lambda\rho\alpha\beta\nu}(p_1,p_2,-k)$}} 
\end{picture}
\\[15mm] 
\begin{picture}(140,40)(-70,-25)
\Photon(-25,0)(0,0)24
\ArrowLine(-20,5)(-10,5)
\Text(-27,0)[r]{$W^+_\alpha(p_1)$}
\Photon(25,0)(0,0){-2}4
\ArrowLine(10,5)(20,5)
\Text(27,0)[l]{$W^+_\beta(p_2)$}
\Photon(0,25)(0,0)24
\ArrowLine(5,20)(5,10)
\Text(0,27)[b]{$Z_\mu(p)$}
\Text(0,-20)[]{\framebox{$-ig\cos\theta_W
    \Ntil_{\alpha\beta\mu}(p_1,p_2)$}} 
\end{picture}
\begin{picture}(140,40)(-70,-25)
\Photon(-25,0)(0,0)24
\ArrowLine(-20,5)(-10,5)
\Text(-27,0)[r]{$W^+_\alpha(p_1)$}
\Photon(25,0)(0,0){-2}4
\ArrowLine(10,5)(20,5)
\Text(27,0)[l]{$W^+_\beta(p_2)$}
\Photon(-25,25)(0,0)24
\ArrowLine(-15,20)(-10,15)
\Text(-25,27)[b]{$Z_\mu(p)$}
\Graviton(25,25)(0,0)24
\Text(25,27)[b]{$h_{\lambda\rho}(q)$}
\ArrowLine(10,15)(15,20)
\Text(0,-20)[]{\framebox{$-i\kappa g\cos\theta_W 
  N_{\lambda\rho\alpha\beta\mu}(p_1,p_2,p)$}} 
\end{picture}
\\[15mm] 
\begin{picture}(140,40)(-70,-25)
\Photon(-25,0)(0,0)24
\Text(-27,0)[r]{$W^+_\alpha$}
\Photon(25,0)(0,0){-2}4
\Text(27,0)[l]{$W^+_\beta(p_2)$}
\Photon(-25,25)(0,0)24
\Text(-25,27)[b]{$Z_\mu$}
\Photon(25,25)(0,0)24
\Text(25,27)[b]{$A_\nu$}
\Text(0,-20)[]{\framebox{$-ieg\cos\theta_W 
  R_{\alpha\beta\mu\nu}$}} 
\end{picture}
\begin{picture}(140,40)(-70,-25)
\Photon(-25,0)(0,0)24
\Text(-27,0)[r]{$W^+_\alpha$}
\Photon(25,0)(0,0){-2}4
\Text(27,0)[l]{$W^+_\beta(p_2)$}
\Photon(-25,25)(0,0)24
\Text(-25,27)[b]{$Z_\mu$}
\Photon(25,25)(0,0)24
\Text(25,27)[b]{$A_\nu$}
\Graviton(0,25)(0,0)24
\Text(0,27)[b]{$h_{\lambda\rho}$}
\Text(0,-20)[]{\framebox{$-i\kappa eg\cos\theta_W 
  R_{\lambda\rho\alpha\beta\mu\nu}$}} 
\end{picture}
\end{center}

\caption{Notations for Feynman rules involving the $W$-bosons in the
  unitary gauge and their extensions to include linearized graviton
  couplings.  For the free $W$-line, the notation represents the
  inverse propagator.}\label{f:Wrules}
\end{figure}

The one-loop diagrams involving the $W$-boson
are shown in \fig{f:Wdiags}, and the relevant Feynman rules are
summarized in \fig{f:Wrules}.  As stated in the Introduction, we employ
the unitary gauge, so that the propagator and gravitational vertex
function of the $W$ are given by
\begin{eqnarray}
\label{DW}
D^{\alpha\beta}_W (k) & = & {1 \over k^2-M_W^2} \left(
-\eta^{\alpha\beta} + {k^\alpha k^\beta \over M_W^2} \right) \,,\\
\label{cW}
c^W_{\lambda\rho\alpha\beta} (k,k') & = & c_{\lambda\rho\alpha\beta}
(k,k') - M_W^2 a'_{\lambda\rho\alpha\beta} \,,
\end{eqnarray}
in analogy with the corresponding quantities for the $Z$ boson.
The trilinear boson couplings involving the $W$
have the following form in the momentum space,
\begin{eqnarray}
\lag_{\rm cubic} = \null - g W_\alpha W_\beta^\dagger W^0_\gamma
N^{\alpha\beta\gamma} (p_1,p_2,p_3) \,,
\end{eqnarray}
where $W_\alpha$ is the field which annihilates the $W^+$ boson, and
$W^0_\gamma$ is the field operator for the neutral SU(2) gauge boson.
The momenta have been written in the same order as the gauge bosons fields,
with $p_2$ flowing out of the vertex and the other two flowing in, and
\begin{eqnarray}
N_{\alpha\beta\gamma} (p_1,p_2,p_3) &=& 
\eta_{\beta\gamma} \eta_{\alpha\sigma} (p_2+p_3)^\sigma 
\null + 
\eta_{\gamma\alpha} \eta_{\beta\sigma} (p_1-p_3)^\sigma 
- \eta_{\alpha\beta} \eta_{\gamma\sigma} (p_1+p_2)^\sigma \,. 
\label{N}
\end{eqnarray}
Since only two of the three momenta are independent in this vertex,
we will often use the shorter notation
\begin{eqnarray}
\label{Ntilde}
\Ntil_{\alpha\beta\gamma} (p_1,p_2) &\equiv& N_{\alpha\beta\gamma}
(p_1,p_2,p_2-p_1) \,.
\end{eqnarray}
Similarly, the quartic $WW\gamma Z$ interaction
in the flat space Lagrangian is given by
\begin{eqnarray}
\lag_{\rm quartic} = - eg \cos\theta_W W_\alpha W_\beta^\dagger Z_\mu
A_\nu R_{\alpha\beta\mu\nu}
\end{eqnarray}
where
\begin{eqnarray}
R_{\alpha\beta\mu\nu} = 2 \eta_{\alpha\beta} 
\eta_{\mu\nu} - \eta_{\alpha\beta\mid\mu\nu} \,.
\label{R4}
\end{eqnarray}
Following the linear approximation to the gravitational interactions
already outlined, we then obtain the corresponding vertices involving
the graviton, characterized by the vertex functions
\begin{eqnarray}
\label{N5}
N_{\lambda\rho\alpha\beta\gamma} (p_1,p_2,p_3) &=&
\eta_{\lambda\rho} 
N_{\alpha\beta\gamma} (p_1,p_2,p_3) 
- \eta_{\lambda\rho\mid\beta\gamma,\alpha\sigma} (p_2+p_3)^\sigma 
\nonumber\\* && 
\null - \eta_{\lambda\rho\mid\gamma\alpha,\beta\sigma} (p_1-p_3)^\sigma 
+ \eta_{\lambda\rho\mid\alpha\beta,\gamma\sigma} (p_1+p_2)^\sigma 
\,,\\
\label{R6}
R_{\lambda\rho\alpha\beta\mu\nu} & = & \eta_{\lambda\rho}
R_{\alpha\beta\mu\nu} 
- 2 \eta_{\lambda\rho\mid\alpha\beta,\mu\nu} + 
\eta_{\lambda\rho\mid\alpha\mu,\beta\nu} + 
\eta_{\lambda\rho\mid\alpha\nu,\beta\mu} \,,
\end{eqnarray}
as indicated in \fig{f:Wrules},
where the symbol $\eta$ with six indices was defined in \Eq{eta6}.
%

\subsection{The amplitude}
Using the Feynman rules just discussed, and defining $\T(W/0)$ by
\begin{eqnarray}
\T(W/1) = \kappa eg \cos\theta_W \T(W/0) \,, 
\label{Tgauge}
\end{eqnarray}
the contributions to $\T(W/0)$ from the diagrams in \fig{f:Wdiags} are
given by
\begin{eqnarray}
i\T(W/a) &=& \lint l \Ntil_{\alpha\beta\mu} (l,l+p)
D_W^{\alpha\tau} (l) \nonumber\\*  && \null \times 
c^W_{\lambda\rho\sigma\tau} (l+q,l) D_W^{\sigma\delta} (l+q)
\Ntil_{\gamma\delta\nu} (l+p,l+q) D_W^{\beta\gamma} (l+p) 
\,, \nonumber\\ 
i\T(W/b) &=& \lint l \Ntil_{\alpha\beta\mu} (l-k,l+q)
D_W^{\alpha\tau} (l-k) 
\nonumber\\*  && \null \times 
\Ntil_{\sigma\tau\nu} (l,l-k) D_W^{\sigma\delta} 
(l) c^W_{\lambda\rho\gamma\delta} (l+q,l) D_W^{\beta\gamma} (l+q) \,,
\nonumber\\  
i\T(W/c) &=& \lint l \Ntil_{\alpha\beta\mu} (l,l+p)
D_W^{\alpha\tau} (l) N_{\lambda\rho\sigma\tau\nu} (l+p,l,-k) 
D_W^{\beta\sigma} (l+p) \,,
\nonumber\\ 
i\T(W/d) &=& \lint l N_{\lambda\rho\alpha\beta\mu} (l,l+k,p)
D_W^{\alpha\tau} (l) \Ntil_{\sigma\tau\nu} (l+k,l)
D_W^{\beta\sigma} (l+k) \,, \nonumber\\  
i\T(W/e) &=& c_{\lambda\rho\delta\nu} (p,k)
D^{\gamma\delta} (p) 
\nonumber\\*  && \null \times 
\lint l \Ntil_{\alpha\beta\mu}(l,l+p)
D_W^{\alpha\tau} (l) 
\Ntil_{\sigma\tau\gamma}(l+p,l) D_W^{\beta\sigma} (l+p) \,,
\nonumber\\  
i\T(W/f) &=& c^Z_{\lambda\rho\mu\alpha} (p,k) D_Z^{\alpha\beta} (k)
  \nonumber\\* 
&& \times \lint l \Ntil_{\gamma\delta\beta} (l,l+k)
  D_W^{\gamma\tau} (l) \Ntil_{\sigma\tau\nu} (l+k,l)
  D_W^{\delta\sigma}(l+k) \,,
  \nonumber\\  
i\T(W/g) &=& R_{\alpha\beta\mu\nu} \lint l 
D_W^{\alpha\tau} (l) c^W_{\lambda\rho\sigma\tau} (l+q,l)
D_W^{\beta\sigma} (l+q) \,, \nonumber\\  
i\T(W/h) &=& R_{\lambda\rho\alpha\beta\mu\nu} \lint l
D_W^{\alpha\beta} (l) \,, \nonumber\\ 
i\T(W/i) &=& c_{\lambda\rho\alpha\nu} (p,k)
D^{\alpha\beta} (p) R_{\sigma\tau\mu\beta} \lint l 
D_W^{\sigma\tau} (l) \,, \nonumber\\  
i\T(W/j) &=& c^Z_{\lambda\rho\mu\alpha} (p,k) D^{\alpha\beta}_Z (k)
R_{\sigma\tau\beta\nu} \lint l 
D_W^{\sigma\tau} (l) \,.
\label{T/W}
\end{eqnarray}
%

\subsection{The electromagnetic transversality condition}
First, using \Eq{transvC} it is immediately seen that
\begin{eqnarray}
k^\nu \T(W/e) = k^\nu \T(W/i) = 0 \,.
\end{eqnarray}
For the rest, we use the Ward identities that have been given in
Appendix \ref{s:WT}. Thus, using \Eq{N5DND}
\begin{eqnarray}
ik^\nu \T(W/d) &=& -k^\nu R_{\lambda\rho\alpha\beta\mu\nu} \lint l 
  D_W^{\alpha\beta} (l) \,,
\end{eqnarray}
and comparing it with the expression for $\T(W/h)$ in \Eq{T/W} we conclude
that
\begin{eqnarray}
k^\nu \T(W/d+h) = 0 \,.
\label{d+g}
\end{eqnarray}
Similarly, applying \Eq{DND}, shifting the momentum in one of the
resulting terms and then applying \Eq{N-N}, we find
\begin{eqnarray}
k^\nu \T(W/f+j) = 0 \,.
\label{f+j}
\end{eqnarray}

Applying \Eq{DND} to $\T(W/a)$ and $\T(W/b)$ we obtain
\begin{eqnarray}
ik^\nu \T(W/a+b) &=& \lint l \Ntil_{\alpha\beta\mu} (l-k,l+q) 
\Big( D_W^{\alpha\delta} (l-k) - D_W^{\alpha\delta} (l) \Big) 
c^W_{\lambda\rho\gamma\delta} (l+q,l) D_W^{\beta\gamma} (l+q)
\nonumber\\ 
&& + \lint l \Ntil_{\alpha\beta\mu} (l,l+p) D_W^{\alpha\tau} (l) 
c^W_{\lambda\rho\sigma\tau} (l+q,l) \Big( D_W^{\beta\sigma} (l+q) - 
D_W^{\beta\sigma} (l+p) \Big) 
\nonumber\\ 
&=& \lint l \Ntil_{\alpha\beta\mu} (l,l+p) D_W^{\alpha\delta} (l) 
\Big( 
c^W_{\lambda\rho\gamma\delta} (l+p,l+k) -
c^W_{\lambda\rho\gamma\delta} (l+q,l) \Big) 
D_W^{\beta\gamma} (l+p)
\nonumber\\ 
&& + \lint l \Big( \Ntil_{\alpha\beta\mu} (l,l+p) - \widetilde
N_{\alpha\beta\mu} (l-k,l+q) \Big) D_W^{\alpha\tau} (l) 
c^W_{\lambda\rho\sigma\tau} (l+q,l) D_W^{\beta\sigma} (l+q) \,.
\nonumber\\* 
\end{eqnarray}
From this, and using \Eqs{C-C}{N-N}, it follows that
\begin{eqnarray}
k^\nu\T(W/a+b+c+g) = 0\,,
\end{eqnarray}
which complemented by \Eqs{d+g}{f+j} establishes the property
\begin{equation}
k^\nu \T(W/0) = 0\,.
\end{equation}
%

\subsection{The gravitational transversality condition}
In similar fashion, here we establish that $\T(W/0)$ satisfies
\begin{equation}
\label{gravtranscond}
q^\lambda\varepsilon^{\nu\ast}(k)\varepsilon^\mu_Z(p)\T(W/0)  = 0\,.
\end{equation}
In the formulas that we obtain below for the contraction of $q$ with
the various amplitudes $\T(W/x)$, we omit writing the polarization
vectors of the photon and the $Z$ to simplify the notation, but it
should be understood that the relations are valid in general only when
the contractions with the polarization vectors as indicated in
\Eq{gravtranscond} are taken, and the on-shell conditions for the
photon and the $Z$ are imposed.

We start with the identity involving the $W$ gravitational vertex and
the propagator in the unitary gauge,
\begin{eqnarray}
D_W^{\alpha\alpha'} (l+q) q^\lambda c^W_{\lambda\rho\alpha\beta}
(l+q,l) D_W^{\beta\beta'} (l) 
&=& l_\rho D_W^{\alpha'\beta'} (l) - (l_\rho + q_\rho) D_W^{\alpha'\beta'}
(l+q) \nonumber\\* 
&& + \eta_{\rho\sigma} \Big( q^{\alpha'} D_W^{\beta'\sigma} (l) +
q^{\beta'} D_W^{\alpha'\sigma} (l+q) \Big) \,,
\label{DqCD}
\end{eqnarray}
which resembles \Eq{SqVS} for fermions.  Using \Eq{DqCD}, we obtain
\begin{eqnarray}
iq^\lambda \T(W/g) &=&  R_{\alpha\beta\mu\nu} \lint l \bigg( 
l_\rho D_W^{\alpha\beta} (l) - (l_\rho+q_\rho) D_W^{\alpha\beta} (l+q) 
\nonumber\\* 
&& +
\eta_{\rho\sigma} q^\alpha D_W^{\beta\sigma} (l+q) + \eta_{\rho\tau}
q^\beta D_W^{\alpha\tau} (l) 
\bigg) \,.
\end{eqnarray}
The first two terms cancel, as can be shown by making a change of variables in
the second one. Using \Eq{R4} the remaining two terms can be written as
\begin{eqnarray}
iq^\lambda \T(W/g) &=& \lint l \bigg( 4 \eta_{\mu\nu} q^\alpha
D^W_{\alpha\rho} (l) - 2 
q_\mu D^W_{\nu\rho} (l) - 2 q_\nu D^W_{\mu\rho} (l) \bigg) \,.
\label{q.Wg}
\end{eqnarray}
Next, using the definition in \Eq{R6}, it is straight forward to write
\begin{eqnarray}
iq^\lambda \T(W/h) = \lint l \bigg( q_\rho R_{\alpha\beta\mu\nu}
D_W^{\alpha\beta} (l) - 2 \Big( \eta_{\nu\rho} q_\mu + \eta_{\mu\rho}
q_\nu \Big) \eta_{\alpha\beta} D_W^{\alpha\beta} (l) - 4 \eta_{\mu\nu}
q^\alpha D^W_{\alpha\rho} (l) \nonumber\\* 
+ 2 q^\alpha \Big( \eta_{\nu\rho} D^W_{\alpha\mu} (l) + \eta_{\mu\rho}
D^W_{\alpha\nu} (l) \Big) + 2 q_\mu D^W_{\nu\rho} (l) + 2 q_\nu
D^W_{\mu\rho} (l) \bigg) \,.
\label{q.Wh}
\end{eqnarray}
Further, using \Eq{q.C}, we obtain
\begin{eqnarray}
iq^\lambda \T(W/i) = 
\bigg[ k_\rho R_{\sigma\tau\mu\nu} - \eta_{\nu\rho} k^\beta
R_{\sigma\tau\mu\beta} - (q_\nu k_\rho -
\eta_{\nu\rho} k \cdot q) {p^\beta \over p^2} R_{\sigma\tau\mu\beta}
\bigg] \lint l D_W^{\sigma\tau} (l) \,.
\label{q.Wi}
\end{eqnarray}
Similarly, using \Eq{q.CZ.D}, we can write
\begin{eqnarray}
iq^\lambda \T(W/j) = \bigg[ \eta_{\mu\rho} q^\alpha
  R_{\sigma\tau\alpha\nu} 
- p_\rho R_{\sigma\tau\mu\nu} \bigg] \lint l D_W^{\sigma\tau} (l) \,,
\label{q.Wj}
\end{eqnarray}
and summing up Eqs.\  (\ref{q.Wg}), (\ref{q.Wh}), (\ref{q.Wi})
and (\ref{q.Wj}), we obtain
\begin{eqnarray}
iq^\lambda \T(W/g+h+i+j) = 2 \eta_{\rho\nu} p^\alpha \lint l
D^W_{\alpha\mu} (l) - (q_\nu k_\rho -
\eta_{\nu\rho} k \cdot q) {p^\beta \over p^2} R_{\sigma\tau\mu\beta}
\lint l D_W^{\sigma\tau} (l) \,.
\label{q.ghij}
\end{eqnarray}

Regarding $\T(W/c)$, by means of \Eq{q.N5RN} we can write
\begin{eqnarray}
iq^\lambda \T(W/c) &=& \lint l N_{\alpha\beta\mu} (l,l+p)
D_W^{\alpha\tau} (l) D_W^{\beta\sigma} (l+p) \bigg[ -k_\rho
  N_{\sigma\tau\nu} (l+p,l) 
\nonumber\\*  && 
+ l_\rho q^\delta R_{\sigma\tau\nu\delta} 
+ p_\rho  N_{\sigma\tau\nu} (l+k,l)
-\eta_{\rho\sigma} q^\delta N_{\delta\tau\nu} (l+k,l) 
\nonumber\\*  && 
-\eta_{\rho\tau} q^\delta N_{\sigma\delta\nu} (l+p,l+q) 
-\eta_{\rho\nu} q^\delta N_{\sigma\tau\delta} (l+p,l) 
\bigg] \,,
\label{q.c}
\end{eqnarray}
and \Eq{q.C} allows us to write the result for $q^\mu\T(W/e)$
as the sum of the two terms,
\begin{eqnarray}
iq^\lambda \T(W/e1) &=& \Big( k_\rho \eta^\delta_\nu 
- k^\delta \eta_{\rho\nu} \Big) \lint l 
N_{\sigma\tau\delta} (l+p,l) N_{\alpha\beta\mu} (l,l+p) 
D_W^{\alpha\tau} (l) D_W^{\beta\sigma} (l+p) 
\,, \nonumber\\ 
iq^\lambda \T(W/e2) &=& - (q_\nu k_\rho -
\eta_{\nu\rho} k \cdot q) {p^\gamma \over p^2} 
\lint l 
N_{\alpha\beta\mu} (l,l+p) N_{\sigma\tau\gamma} (l+p,l)
D_W^{\alpha\tau} (l) D_W^{\beta\sigma} (l+p) \nonumber\\ 
&=&  (q_\nu k_\rho - \eta_{\nu\rho} k \cdot q) 
{p^\delta \over p^2} R_{\sigma\tau\mu\delta}
\lint l D_W^{\sigma\tau} (l) \,,
\end{eqnarray}
where we have applied \Eq{NDND} in the last step.  If we consider the
sum $q^\mu\T(W/c+e1)$, the term from $q^\mu\T(W/e1)$ containing
$k_\rho$ cancels an identical term from \Eq{q.c}.  The remaining term
from $q^\mu\T(W/e1)$ combines with the term proportional to
$\eta_{\rho\nu}$ from \Eq{q.c}, and applying \Eq{NDND} they yield
\begin{eqnarray}
\eta_{\rho\nu} p^\delta R_{\sigma\tau\mu\delta} \lint l
D_W^{\sigma\tau} (l) \,.
\end{eqnarray}
Substituting the expression for $R_{\alpha\beta\mu\delta}$ given in
\Eq{R4}, the term proportional $p^\mu$ vanishes due to \Eq{e.p=0} and
the remainder cancels one of the terms of \Eq{q.ghij}. The other term
of \Eq{q.ghij} is canceled by the $q^\mu\T(W/e2)$ contribution, and in
this way we obtain
\begin{eqnarray}
iq^\lambda \T(W/c+e+g+h+i+j) &=& 
\lint l N_{\alpha\beta\mu} (l,l+p) D_W^{\alpha\tau}
(l) D_W^{\beta\sigma} (l+p) \bigg[ 
l_\rho q^\delta  R_{\sigma\tau\nu\delta} 
\nonumber\\*  && 
+ p_\rho  N_{\sigma\tau\nu} (l+k,l)
-\eta_{\rho\sigma} q^\delta N_{\delta\tau\nu} (l+k,l) 
\nonumber\\*  && 
-\eta_{\rho\tau} q^\delta N_{\sigma\delta\nu} (l+p,l+q) 
\bigg] \,.
\label{q.ceghij}
\end{eqnarray}

For the contraction of $\T(W/f)$ we use \Eq{q.CZ.D}
to write it in the form
\begin{eqnarray}
iq^\lambda \T(W/f) = \Big( \eta_{\rho\mu} q^\delta - p_\rho
  \eta_\mu^\delta \Big) \lint l 
  N_{\alpha\beta\delta} (l,l+k) N_{\sigma\tau\nu} (l+k,l)
  D_W^{\alpha\tau} (l) D_W^{\beta\sigma} (l+k) \,,
\end{eqnarray}
and applying the identity of \Eq{q.N5NN}
\begin{eqnarray}
iq^\lambda \T(W/d) &=& \lint l \Ntil_{\sigma\tau\nu} (l+k,l)
D_W^{\alpha\tau} (l) D_W^{\beta\sigma} (l+k) \bigg[ (p_\rho - l_\rho)
  \Ntil_{\alpha\beta\mu} (l,l+k) \nonumber\\*  && 
+ l_\rho \Ntil_{\alpha\beta\mu} (l-q,l-q+k)
- k_\rho \Ntil_{\alpha\beta\mu} (l,l+p)
-\eta_{\rho\alpha} q^\delta N_{\delta\beta\mu} (l-q,l+k) 
\nonumber\\*  && 
-\eta_{\rho\beta} q^\delta N_{\alpha\delta\mu} (l,l+p) 
-\eta_{\rho\mu} q^\delta N_{\alpha\beta\delta} (l,l+k) \bigg] \,.
\end{eqnarray}
Adding these results and using \Eq{Nab-Nab'},
\begin{eqnarray}
iq^\lambda \T(W/d+f) &=& \lint l N_{\sigma\tau\nu} (l+k,l)
D_W^{\alpha\tau} (l) D_W^{\beta\sigma} (l+k) \bigg[ 
l_\rho q^\delta R_{\alpha\beta\mu\delta} 
\nonumber\\*  && 
-k_\rho N_{\alpha\beta\mu} (l,l+p) 
-\eta_{\rho\alpha} q^\delta N_{\delta\beta\mu} (l-q,l+k) 
-\eta_{\rho\beta} q^\delta N_{\alpha\delta\mu} (l,l+p) 
 \bigg] \,.
\label{q.df}
\end{eqnarray}

Finally, for $\T(W/a)$ and $\T(W/b)$ we use \Eq{DqCD} once again.
By redefining the integration variable in
certain terms, the results can be written as
\begin{eqnarray}
iq^\lambda \T(W/a) &=& \lint l \bigg[ N_{\alpha\beta\mu} (l,l+p)
N_{\sigma\tau\nu} (l+p,l+q) D_W^{\beta\sigma} (l+p) 
\Big( l_\rho D_W^{\alpha\tau} (l) + \eta^{\alpha\delta} q^\tau
D^W_{\delta\rho} (l) \Big) 
\nonumber\\*  && 
+ N_{\alpha\beta\mu} (l-q,l+k) N_{\sigma\tau\nu} (l+k,l)
 D_W^{\beta\sigma} (l+k) \Big( -l_\rho D_W^{\alpha\tau} (l) +
 \eta^{\tau\delta} q^\alpha D^W_{\delta\rho} (l) \Big) \bigg] \,, 
\label{q.a} 
\\ 
iq^\lambda \T(W/b) &=& \lint l N_{\alpha\beta\mu} (l,l+p)
N_{\sigma\tau\nu} (l+k,l) D_W^{\alpha\tau} (l) 
\bigg[ (l_\rho+k_\rho) D_W^{\beta\sigma} (l+k)
\nonumber\\*  && 
  - (l_\rho + p_\rho) D_W^{\beta\sigma} (l+p) 
+ \eta^{\beta\delta} q^\sigma D^W_{\delta\rho} (l+p) 
+ \eta^{\sigma\delta} q^\beta D^W_{\delta\rho} (l+k) \bigg] \,.
\label{q.b}
\end{eqnarray}
Adding Eqs.\ (\ref{q.ceghij}), (\ref{q.df}), (\ref{q.a}) and
(\ref{q.b}), and applying the identities of \Eqs{Nab-Nab'}{Nab-Na'b},
we verify that all the terms cancel, which proves \Eq{gravtranscond}.

\subsection{Calculation of the form factors}
As we have already argued, the virtue of having proved that $\T(W/0)$
satisfies the electromagnetic and gravitational transversality
conditions is that we are now assured that it has the structure given
in \Eq{Tgeneral}. Furthermore, there cannot be any contribution to the
factors $F_1$ and $F_2$ since the Feynman rules do not involve the
Levi-Civita tensor. Therefore, to calculate the contribution to the
form factor $F$, we can just look at the $kkqq$ terms, as we did in
the fermion case.  Fortunately, as we now show, only $\T(W/a)$ and
$\T(W/b)$ produce such terms, and they are in fact equal so we need to
evaluate only one of them.

Since $R_{\alpha\beta\mu\delta}$ is independent of momentum,
the amplitude of $\T(W/g)$ does not depend on $k$, and
consequently does not produce a $kkqq$ term.  The same argument applies
to the amplitude $\T(W/j)$, which is independent of $k$ as well as $q$.
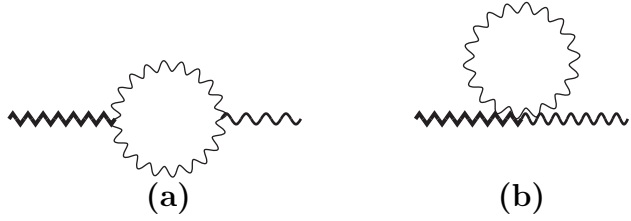
\begin{figure}
\begin{center}
\begin{picture}(130,80)(-70,-40)
\PhotonArc(0,0)(20,0,360)2{20}
\SetWidth{1.5} 
\ZigZag(-60,0)(-20,0)27
\SetWidth{1}
\Photon(20,0)(50,0)24
\Text(0,-30)[]{\large\bf(a)}
\end{picture}
\begin{picture}(130,80)(-70,-40)
\PhotonArc(0,22)(20,0,360)2{18}
\SetWidth{1.5} 
\ZigZag(-40,0)(0,0)27
\SetWidth{1}
\Photon(0,0)(40,0)27
\Text(0,-30)[]{\large\bf(b)}
\end{picture}
\end{center}

\caption{One-loop diagrams for the $\gamma Z$ polarization mixing
tensor involving internal $W$-boson lines.}\label{f:PiWloop}
\end{figure}
Next, consider the amplitudes $\T(W/f)$ and $\T(W/j)$. Their sum
can be written in the form
\begin{eqnarray}
\T(W/f) + \T(W/j) = c^Z_{\lambda\rho\mu\alpha} (p,k) D^{\alpha\beta}_Z (k)
  \Pi^{(W)}_{\beta\nu}(k) \,,
\end{eqnarray}
where $\Pi^{(W)}_{\mu\nu}$ denotes the contribution of $W$-loops to
the $\gamma Z$ polarization mixing tensor, represented by the diagrams
shown in \fig{f:PiWloop}.  This tensor, as can be readily verified
explicitly, is transverse and therefore it has the same form given in
\Eq{Piform}.  We can then apply the same argument given in Section
\ref{subsec:fermionscalc} to dismiss the $\T(f/e)$ and $\T(f/f)$
amplitudes in the fermion case, to conclude that the combination
$\T(W/f) + \T(W/j)$ does not give any $kkqq$ term, and a similar
argument holds for $\T(W/e) + \T(W/i)$ as well.

Considering $\T(W/d)$, the $q$-dependence can come only from the
factor $N_{\lambda\rho\alpha\beta\mu} (l,l+k,p)$ in the integrand
after we substitute $p=k+q$.  Since
\begin{eqnarray}
N_{\lambda\rho\alpha\beta\mu} (l,l+k,k+q) = \Big( 
\eta_{\lambda\rho\mid\mu\alpha,\beta\sigma} 
- \eta_{\lambda\rho\mid\beta\mu,\alpha\sigma} \Big) q^\sigma 
+ \mbox{terms independent of $q$}\,,
\end{eqnarray}
we obtain
\begin{eqnarray}
i\T(W/d) = \Big( 
\eta_{\lambda\rho\mid\mu\alpha,\beta\sigma} 
- \eta_{\lambda\rho\mid\beta\mu,\alpha\sigma} \Big) q^\sigma 
\lint l 
D_W^{\alpha\tau} (l) \Ntil_{\sigma\tau\nu} (l+k,l)
D_W^{\beta\sigma} (l+k) + \cdots \,,
\end{eqnarray}
where the dots denote $q$-independent terms and therefore do not
contain any $kkqq$ terms.  The remaining integral in this equation
yields a three-index tensor that depends only on $k$. Since $k^2 =
0$, the only  contribution from the integral that yields a term
with four powers of momenta in $\T(W/d)$ contains the factor
$k^\alpha k^\beta k_\nu$. \Eq{e.k=0} then implies that it does not
contribute to the amplitude,
and thus we conclude that $\T(W/d)$ does not yield any $kkqq$ term.
The same conclusion is reached for $\T(W/c)$ as well,
for which it is only necessary to note that
\begin{eqnarray}
\T(W/d) (p,k) = T^{\di(W/c)}_{\lambda\rho\nu\mu}(-k, -p)\,,
\end{eqnarray}
where we have explicitly indicated the $(p,k)$ dependence in order to
state the argument. 

Thus, only $\T(W/a)$ and $\T(W/b)$ can give the required type of terms.
Furthermore, by  changing the integration variable from
$l$ to $-l+k$ in the expression for $\T(W/b)$
given in \Eq{T/W} and using the symmetry properties of the couplings
involved, it follows that
\begin{eqnarray}
\T(W/a) = \T(W/b) \,.
\end{eqnarray}
Therefore we need to evaluate only one of these two, and we choose
$\T(W/b)$. We first write the amplitude it in the form
\begin{eqnarray}
i\T(W/b) &=& \lint l 
\Delta_W (l-k) \Delta_W (l) \Delta_W (l+q) \S.. \,,
\label{TS}
\end{eqnarray}
where we have defined
\begin{eqnarray}
\Delta_W (l) = {1 \over l^2 - M_W^2} \,,
\end{eqnarray}
and
\begin{eqnarray}
\S.. &=& \Ntil_{\alpha\beta\mu} (l-k,l+q)
\Ntil_{\sigma\tau\nu} (l,l-k) 
c^W_{\lambda\rho\gamma\delta} (l+q,l) 
\nonumber\\*  && \null \times 
\left[ - \eta^{\alpha\tau} +
  {(l-k)^\alpha (l-k)^\tau \over M_W^2} \right]
\left[ - \eta^{\sigma\delta} +
  {l^\sigma l^\delta \over M_W^2} \right]
\left[ - \eta^{\beta\gamma} +
  {(l+q)^\beta (l+q)^\gamma \over M_W^2} \right] 
\nonumber\\ 
& \equiv & \S..^{(0)} + \S..^{(2)} + \S..^{(4)} + \S..^{(6)} \,.
\label{S}
\end{eqnarray}
In the last step, we divide the terms in $\S..$ into four terms, each one
labeled by an index that denotes the number of inverse powers of $M_W$
that it contains.

We consider first the term which has an overall factor of $1/M_W^6$,
\begin{eqnarray}
\S..^{(6)} & = & {1 \over M_W^6} (l-k)^\alpha (l-k)^\tau l^\sigma
l^\delta (l+q)^\beta (l+q)^\gamma \nonumber\\
& & \times
\Ntil_{\alpha\beta\mu} (l-k,l+q)
\Ntil_{\sigma\tau\nu} (l,l-k) c^W_{\lambda\rho\gamma\delta}(l+q,l) \,. 
\label{S6}
\end{eqnarray}
To reduce this term and the other ones, we use the following identities 
\begin{eqnarray}
a^\alpha \Ntil_{\alpha\beta\gamma} (a,b) &=&
P_{\beta\gamma} (b-a) - P_{\beta\gamma} (b) \,, \nonumber \\ 
b^\beta \Ntil_{\alpha\beta\gamma} (a,b) &=&
P_{\gamma\alpha} (b-a) - P_{\gamma\alpha} (a) \,, \nonumber \\ 
a^\alpha b^\beta \Ntil_{\alpha\beta\gamma} (a,b) &=&
a \cdot (b-a) (b-a)_\gamma - (b-a)^2 a_\gamma \,,
\label{Ncontract}
\end{eqnarray}
where
\begin{eqnarray}
P_{\alpha\beta}(l) = - l^2 \eta_{\alpha\beta} + l_\alpha l_\beta \,.
\label{P}
\end{eqnarray}
The identities in \Eq{Ncontract}, which are similar to the relations
given in \Eq{D-D}, follow directly from \Eqs{N}{Ntilde}.
Applying them to the particular combination that appears in \Eq{S6}, we obtain
\begin{eqnarray}
l^\sigma (l-k)^\tau \Ntil_{\sigma\tau\nu} (l,l-k) = l \cdot k
k_\nu - k^2 l_\nu \,,
\label{llN}
\end{eqnarray}
which shows that $\S..^{(6)}$ does not contribute to the amplitude after
using the on-shell conditions for the photon given in \Eqs{e.k=0}{k^2=0}.

$\S..^{(4)}$ contains three terms, with an overall
factor of $1/M_W^4$.  One of them vanishes due to \Eq{llN}.  For the
others, we use the definition of \Eq{cW} and the
identity of \Eq{transvC} to obtain another useful identity
\begin{eqnarray}
l'^\tau c^W_{\lambda\rho\sigma\tau} (l,l') =  - M_W^2 l'^\tau
a'_{\lambda\rho\sigma\tau}\,,
\end{eqnarray}
which allows us to write
\begin{eqnarray}
\S..^{(4)} &=& {1 \over M_W^2} 
a'_{\lambda\rho\gamma\delta} 
\Ntil_{\alpha\beta\mu} (l-k,l+q)
\Ntil_{\sigma\tau\nu} (l,l-k) 
\nonumber\\*  && \null \times 
\left[ \eta^{\sigma\delta}
  (l-k)^\alpha (l-k)^\tau + \eta^{\alpha\tau}
  l^\sigma l^\delta \right]
  (l+q)^\beta (l+q)^\gamma \,,
\end{eqnarray}
where in the last step we have used the definition of
$c^W_{\lambda\rho\sigma\tau}$ from
\Eq{cW}, along with the identity of \Eq{transvC}. Applying
the identities in \Eq{Ncontract}, and omitting any term that does
not contribute to the amplitude due to the on-shell conditions for
the photon and the $Z$, the expression for $\S..^{(4)}$ reduces to
\begin{eqnarray}
\S..^{(4)} = {1 \over M_W^2} 
a'_{\lambda\rho\gamma\delta} (l+q)^\gamma 
\bigg[ M_Z^2 (l-k)_\mu P^\delta_\nu (l) 
+ l^\delta 
\Big( M_Z^2 \eta_{\alpha\mu} + P_{\alpha\mu} (l-k) \Big)
P^\alpha_\nu (l-k) \bigg] \,.
\label{S4a}
\end{eqnarray}
Using the identity
\begin{eqnarray}
P_{\alpha\mu} (l-k) P^\alpha_\nu (l-k) = - (l-k)^2 P_{\mu\nu} (l-k) \,,
\end{eqnarray}
and noticing that we can make the replacements
\begin{eqnarray}
a'_{\lambda\rho\gamma\delta} (l+q)^\gamma & \rightarrow & \null -
\eta_{\lambda\delta} l_\rho - \eta_{\rho\delta} l_\lambda \,,\nonumber\\
P^\delta_\nu (l) & \rightarrow & l^\delta l_\nu \,,
\label{a'l}
\end{eqnarray}
because the neglected terms give a vanishing contribution
to the amplitude when the on-shell graviton condition is imposed, we
obtain 
\begin{eqnarray}
\S..^{(4)}
&=& \null - {2 \over M_W^2} l_\lambda l_\rho (l-k)_\mu l_\nu \Big[
  2M_Z^2 - (l-k)^2 \Big] + \cdots\,,
\end{eqnarray}
where the dots denote terms which cannot produce any $kkqq$ term.
This can be reduced further by writing
\begin{eqnarray}
(l-k)^2 = \Delta_W^{-1}(l-k) + M_W^2 \,,
\end{eqnarray}
and remembering that $\S..^{(4)}$ is to be substituted in \Eq{TS}.
The $\Delta_W^{-1}(l-k)$ term then cancels with the $\Delta_W (l-k)$
in \Eq{TS}, and the resulting integral, which depends only on $q$,
does not produce a $kkqq$ type term. Thus the relevant part of $\S..^{(4)}$ is
just
\begin{eqnarray}
\S..^{(4)} 
&=& \left(2 - {4M_Z^2 \over M_W^2} \right) l_\lambda l_\rho (l-k)_\mu
l_\nu + \cdots\,.
\label{S4final}
\end{eqnarray}

$\S..^{(2)}$, which from \Eq{S} is given by,
\begin{eqnarray}
\S..^{(2)} &=& {1 \over M_W^2} \Ntil_{\alpha\beta\mu} (l-k,l+q)
\Ntil_{\sigma\tau\nu} (l,l-k) 
c^W_{\lambda\rho\gamma\delta} (l+q,l) 
\nonumber\\*  && \null \times 
\left[ 
\eta^{\alpha\tau} \eta^{\sigma\delta} 
  (l+q)^\beta (l+q)^\gamma + 
\eta^{\alpha\tau} \eta^{\beta\gamma} 
  l^\sigma l^\delta + 
\eta^{\sigma\delta}
\eta^{\beta\gamma} 
  (l-k)^\alpha (l-k)^\tau 
\right] \,,
\end{eqnarray}
can be treated in similar fashion. By
using \Eqss{transvC}{Ncontract}{a'l}, it can be reduced to
\begin{eqnarray}
\S..^{(2)} &=& 
\Big( P^\tau_\mu (p) - P^\tau_\mu (l-k) \Big)
\Ntil_{\sigma\tau\nu} (l,l-k) 
\Big(\eta^\sigma_\lambda l_\rho + \eta^\sigma_\rho l_\lambda \Big)
\nonumber\\*  && 
+ \Ntil_{\alpha\beta\mu} (l-k,l+q)
\Big( P^\alpha_\nu (k) - P^\alpha_\nu (l-k) \Big)
\Big(\eta^\beta_\lambda l_\rho + \eta^\beta_\rho l_\lambda \Big)
\nonumber\\*  && 
+ {1 \over M_W^2} \Big( P^\gamma_\mu (p) - P^\gamma_\mu (l+q) \Big)
\Big( P^\delta_\nu (k) - P^\delta_\nu (l) \Big)
c^W_{\lambda\rho\gamma\delta} (l+q,l) 
+ \cdots \,,
\end{eqnarray}
which can be simplified further by neglecting the terms
that do not contribute to the on-shell amplitude and focusing
on the $kkqq$ type terms. Omitting the rest of those details, we
finally obtain, for the relevant part of $\S..^{(2)}$,
\begin{eqnarray}
\S..^{(2)} &=& 6l_\lambda l_\rho (l-k)_\mu l_\nu \,.
\label{S2final}
\end{eqnarray}
Adding \Eqs{S4final}{S2final} we then obtain
\begin{eqnarray}
\label{S2+4}
\S..^{(2+4)} = 4\left(2 - {M_Z^2 \over M_W^2} \right) l_\lambda l_\rho
(l-k)_\mu l_\nu + \cdots \,,
\end{eqnarray}
and it should be remembered that $\S..^{(6)}$
does not contribute to the $kkqq$ terms, so that it need not be considered.
Substituting \Eq{S2+4} into \Eq{TS} and parameterizing the integral in
the standard way, we obtain the result from this part,
\begin{eqnarray}
\T(W/b) \Bigg|_{(2+4)} = \null - 8i \left(2 - {M_Z^2 \over M_W^2}
  \right) \lintd l \int_0^1 dx
  \int_0^{1-x} dy \; 
  {f_{\lambda\rho\mu\nu}(l+xk-yq) \over \Big[ l^2 - M_W^2 + xy M_Z^2
  \Big]^3} + \cdots \,,
\end{eqnarray}
where now
\begin{eqnarray}
f_{\lambda\rho\mu\nu}(l) = l_\lambda l_\rho (l-k)_\mu l_\nu \,.
\end{eqnarray}
Choosing the $kkqq$ terms as we have indicated earlier,
we obtain the contribution from this part to the $kkqq$ term,
\begin{eqnarray}
\T(W/b) \Bigg|_{(2+4)} &=& 8i \left(2 - {M_Z^2 \over M_W^2}
  \right) \; k_\lambda k_\rho q_\mu q_\nu 
\nonumber\\* && \times
  \lint{l} \int_0^1 dx \int_0^{1-x} dy \; 
  {x^2 y (1-x-y) \over \Big[ l^2 - M_W^2 + xy M_Z^2
  \Big]^3} + \cdots \,.
\label{S2+S4}
\end{eqnarray}
where we have continued the integral to four dimensions since it is finite.

We now work the $\S..^{(0)}$ term in \Eq{S}, which is given by
\begin{eqnarray}
\S..^{(0)} &=& \null - \eta^{\alpha\tau} \eta^{\sigma\delta}
\eta^{\beta\gamma} \Ntil_{\alpha\beta\mu} (l-k,l+q)
\Ntil_{\sigma\tau\nu} (l,l-k) 
c^W_{\lambda\rho\gamma\delta} (l+q,l) \,.
\label{S0}
\end{eqnarray}
From the definition of the cubic couplings in \Eq{N}, and using \Eq{transvC},
the part that can possibly contain $kkqq$ terms is
\begin{eqnarray}
\S..^{(0)} 
&=& \Big[ \eta^\gamma_\mu (l+k+2q)_\alpha - 2
  \eta_{\mu\alpha} (k+q)^\gamma - 2 \eta^\gamma_\alpha (l+q)_\mu \Big]
\nonumber \\* && \times
\Big[ 2\eta^\alpha_\nu k^\delta - \eta_\nu^\delta (l+k)^\alpha + 2
  \eta^{\delta\alpha} l_\nu \Big]
c_{\lambda\rho\gamma\delta} (l+q,l) + \cdots \,,
\label{S0a}
\end{eqnarray}
where we have also made use of \Eqs{e.k=0}{k=-q}.  In addition, we
have noticed that the term proportional to $M_W^2$ in
$c^W_{\lambda\rho\gamma\delta}$ does not produce a term quartic in the
momenta, since the cubic couplings are linear in momenta.  From the
definition in \Eq{c}, and excluding all terms which either vanish
on contraction with the
polarization factors or do not contribute to the $kkqq$ type terms
because they do not have enough factors of uncontracted momenta,
we find that we can substitute
\begin{eqnarray}
c_{\lambda\rho\gamma\delta} (l+q,l) = \Big[ - \eta_{\gamma\delta}
  l_\lambda l_\rho + \eta_{\lambda\gamma} l_\rho (l+q)_\delta +
  \eta_{\lambda\delta} l_\gamma l_\rho \Big] + (\lambda
  \leftrightarrow \rho)\,.
\end{eqnarray}
The rest of the calculation is just straight forward algebra, and the
result obtained is
\begin{eqnarray}
\S..^{(0)} = 16 l_\lambda l_\rho \Big[ (l+q)_\mu l_\nu + q_\mu q_\nu
  \Big]  + 8 (k_\lambda l_\rho + l_\lambda k_\rho) \Big[ l_\mu
  q_\nu - q_\mu l_\nu \Big] + \cdots \,.
\end{eqnarray}
Substituting this expression into \Eq{TS} we then obtain
\begin{eqnarray}
\T(W/b) \Bigg|_{(0)} = 32i \; k_\lambda k_\rho q_\mu q_\nu
  \lint4 l \int_0^1 dx \int_0^{1-x} dy \; 
  {x^2 y (1-x-y) \over \Big[ l^2 - M_W^2 +
  xy M_Z^2 \Big]^3} +
  \cdots \,,
\end{eqnarray}
which has to be added to the contribution obtained in \Eq{S2+S4} to
obtain the complete expression for the $kkqq$ terms in
$\T(W/b)$. Remembering that $\T(W/a)$ and $\T(W/b)$ give identical
contributions, and recalling the overall factors defined in
\Eq{Tgauge}, we obtain the $W$ contribution to the form factor $F$,
\begin{eqnarray}
F^\di(W/0) &=& 
{\kappa eg \cos\theta_W \over 4\pi^2 M^2_Z} \left(6 - {M_Z^2 \over M_W^2}
  \right) \; I(M_W/M_Z) \,,
\label{FW}
\end{eqnarray}
where $I(A)$ has been defined in \Eq{I}.

\section{Decay rate and discussion}\label{s:dk}
The on-shell amplitude is parameterized by the three form factors
defined in \Eq{Tgeneral}. Our calculations in Sec.~\ref{s:fl} and
Sec.~\ref{s:Wl} show, first of all, that
\begin{eqnarray}
F_1 = F_2 = 0 \,,
\label{F1F2}
\end{eqnarray}
which is a consequence of CP invariance.  The terms containing $F_1$
and $F_2$ in \Eq{Tgeneral} contain the Levi-Civita tensor, and are
therefore odd under time reversal and under CP.
In the standard electroweak model, CP violation enters any amplitude
only through the charged current interactions of fermions, which do
not appear in the one-loop amplitudes for the present process.  Hence,
at the one-loop level, the amplitude is CP conserving, and \Eq{F1F2}
reflects that fact.

The only non-vanishing form factor in one-loop is $F$, for which the
results given in \Eqs{Fferm}{FW} are combined to give 
\begin{eqnarray}
F = {\kappa eg \over 4\pi^2 M_Z^2 \cos\theta_W} 
\left[ \cos^2\theta_W \left(6 - {1 \over \cos^2\theta_W} \right)
  I(M_W/M_Z) - 
  2 \sum_f Q_f X_f I(m_f/M_Z) \right] \,.
\label{F}
\end{eqnarray}
where $I$ is the integral defined in \Eq{I}.
The integral cannot be performed analytically, but we can make some
approximations that are sufficient for our purposes.
In the denominator of the integrand
the combination $xy$ has the maximum value $\frac14$
within the range of integration. We thus calculate the integral in
two extreme cases,
\begin{eqnarray}
I(A) = \cases{\null - \displaystyle{1 \over 24} & for $A\ll \frac14$, \cr 
\displaystyle{1 \over 360A^2} & for $A\gg \frac14$.}
\end{eqnarray}
For all the fermions except the top quark, we use the first form, whereas
for the top quark and the $W$ in the loop, we use the second. Thus,
\begin{eqnarray}
F = {\kappa eg \over 4\pi^2 M_Z^2 \cos\theta_W} \times 
\left[ {1\over 360} \left(6 - {1 \over \cos^2\theta_W}
  \right)
+ {5 \over 12} \left(1 - \frac{2 M_Z^2}{15m_t^2}\right) - {10 \over 9}
  \sin^2\theta_W \left( 1 - \frac{M_Z^2}{75 m_t^2}\right) \right] \,,
\label{Fnum}
\end{eqnarray}
where we have used the mass relation $M_W=M_Z\cos\theta_W$.  Using
$e=g\sin\theta_W$ and $\sin^2\theta_W=0.23$, this gives
\begin{eqnarray}
\label{Fsm}
F = 0.4 {\kappa e^2 \over 4\pi^2 M_Z^2} \,.
\end{eqnarray}

The decay rate is determined straightforwardly from
\Eq{Tgeneral}. Using the familiar polarization sum formulas
for the photon and the $Z$, as well as
the corresponding one for the graviton\cite{scadron},
\begin{eqnarray}
\sum_{\rm pol} \varepsilon^{\nu*}(k) \varepsilon^{\nu'}(k) &=& -
\eta^{\nu\nu'} \,,\\ 
\sum_{\rm pol} \varepsilon^\mu_Z (p) \varepsilon^{\mu^\prime}_Z (p) &=& -
\eta^{\mu\mu'} + {p^\mu p^{\mu'} \over M_Z^2} \,, \\ 
\sum_{\rm pol} {\cal E}^{\lambda\rho*}(q) {\cal E}^{\lambda'\rho'}(q)
&=& \frac12 \Big( \eta^{\lambda\lambda'} \eta^{\rho\rho'} +
\eta^{\lambda\rho'} \eta^{\lambda'\rho} -
\eta^{\lambda\rho} \eta^{\lambda'\rho'} 
 \Big) \,,
\end{eqnarray}
we find
\begin{eqnarray}
\Gamma = {M_Z^7 F^2 \over 96 \pi} \,,
\label{Gamma}
\end{eqnarray}
and from \Eq{Fsm}
\begin{eqnarray}
\Gamma = 0.1 {\alpha^2 G M_Z^3 \over \pi^2} \,.
\end{eqnarray}

This result confirms our expectation in \Eq{estimate} about the
smallness of the rate. However, as already mentioned in the
Introduction, one of the primary motivations for performing the
calculation was to understand some of the intricacies involved and
resolve some of the technical complications in a way that can be used
in similar, perhaps more complicated, calculations.  In this sense,
the proof of the general form of the amplitude and its parameterization
given in \Eq{Tgeneral}, together with the all the identities,
manipulations and tricks that we have used both in the explicit
calculation of the form factors as well as in the proofs of the
consistency conditions (the electromagnetic and gravitational
transversality conditions) are useful in their own right,
independently of the fact that we have applied them in the particular
context of the $Z$ decay. It is particularly enlightening the fact
that the use of the unitary gauge did not lead to any of the
inconsistencies that are sometimes attributed to using that gauge. In
fact, as we showed, the diagrams calculated with this gauge give an
amplitude that is consistent with the transversality conditions implied
by the electromagnetic and gravitational gauge invariance, which in
turn allowed us to determine the amplitude and calculate the relevant
form factor in a systematic and consistent fashion.  By having
considered this simpler system, it has allowed us to understand and
develop some techniques that we believe can be useful for considering
more complicated processes.

\subsection*{Acknowledgements}
The work of JFN was supported by the U.S. National Science
Foundation under Grant 0139538.  PBP wants to thank G. Bhattacharyya
and S. Rakshit for discussions.

\appendix
\section*{Appendices}
\section{Ward-Takahashi identities}\label{s:WT}
The simplest Ward-Takahashi (WT) identity familiar to us through QED,
which relates the fermion two-point function to the fermion-photon
vertex.  Diagrammatically, it can be written as
\begin{eqnarray}
-eQ \left[\bigg( \begin{picture}(50,30)(0,0)
\SetWidth{1.5}
\ArrowLine(0,0)(50,0)
\Text(25,-7)[]{$l+r$}
\end{picture} \bigg)
- 
\bigg( \begin{picture}(50,30)(0,0)
\SetWidth{1.5}
\ArrowLine(0,0)(50,0)
\Text(25,-7)[]{$l$}
\end{picture} \bigg) \right]
= r^\nu
\left( \begin{picture}(50,30)(0,0)
\SetWidth{1.5}
\ArrowLine(0,0)(25,0)
\ArrowLine(25,0)(50,0)
\Photon(25,0)(25,25)23
\Text(28,25)[l]{$r^\nu$}
\Text(10,-7)[]{$l+r$}
\Text(40,-7)[]{$l$}
\Vertex(25,0)5
\end{picture} \right) \,,
\label{S-S}
\end{eqnarray}
where the fermion has charge $eQ$.  The tree-level version of this
equality was presented in \Eq{Sk/S}.  After the introduction of
gravity, vertices involving gravitons appear in the theory.  In a
similar fashion, one can now prove the WT identity
\begin{eqnarray}
-eQ \left[
\left( \begin{picture}(50,30)(0,0)
\SetWidth{1.5}
\ArrowLine(0,0)(25,0)
\Text(10,-7)[]{$l+t$}
\ArrowLine(25,0)(50,0)
\Text(40,-7)[]{$l+r$}
\Graviton(25,0)(25,25)23
\Vertex(25,0)5
\end{picture} \right)
- 
\left( \begin{picture}(50,30)(0,0)
\SetWidth{1.5}
\ArrowLine(0,0)(25,0)
\Text(10,-7)[]{$l+s$}
\ArrowLine(25,0)(50,0)
\Text(40,-7)[]{$l$}
\Graviton(25,0)(25,25)23
\Vertex(25,0)5
\end{picture} \right) \right]
= r^\nu
\left( \begin{picture}(50,30)(0,0)
\SetWidth{1.5}
\ArrowLine(0,0)(25,0)
\Text(10,-7)[]{$l+t$}
\ArrowLine(25,0)(50,0)
\Text(40,-7)[]{$l$}
\Photon(25,0)(50,25)23
\Text(45,25)[r]{$r^\nu$}
\Graviton(25,0)(0,25)23
\Vertex(25,0)5
\end{picture} \right) \,.
\label{V-V(fig)}
\end{eqnarray}
In this and other relations in this appendix, we use the shorthand
\begin{eqnarray}
t = r + s \,,
\end{eqnarray}
where $r$ and $s$ are the momenta of the photon and the graviton lines
respectively, both considered outgoing.  Recalling the definition of
the vertices given in \fig{f:frules}, we see that at tree level,
the diagrammatic identity of \Eq{V-V(fig)} implies the relation
\begin{eqnarray}
 V_{\lambda\rho} (l+r+s,l+r) - V_{\lambda\rho} (l+s,l) = \null -
 a_{\lambda\rho\alpha\beta} \gamma^\alpha r^\beta \,,
\label{V-V}
\end{eqnarray}
which can be easily checked from the expressions for these vertices
appearing in \Eqs{V}{a'}.

The general nature of these WT identities are now clear.  For the
charged $W^+$ bosons, we can write similar identities.  Of course,
these results depend on the gauge condition, and involves the
unphysical scalar fields as well as ghost fields.  In the unitary
gauge, however, the unphysical scalar fields and the ghost fields are
not present, and the relation look particularly simple.  Here, we
summarize some relations of this sort.  In the diagrams which appear
within the equations below, the horizontal lines denote $W^+$, where
the left line has incoming momentum and the right line has outgoing.
The $Z$-line appears in a saw-tooth pattern, with an inward momentum
$p$.  The momentum convention for the photon and the graviton lines
have already been stated.

First, we can have the identity from a diagram similar to that in
\Eq{S-S}, with the fermion lines replaced by the $W^+$-lines.  At the
tree level, this will imply
\begin{eqnarray}
D^{-1}_{\alpha\beta} (l+r) - D^{-1}_{\alpha\beta} (l) = r^\gamma 
\Ntil_{\alpha\beta\gamma} (l+r,l) \,.
\label{D-D}
\end{eqnarray}
This can also be written as
\begin{eqnarray}
r^\gamma D^{\alpha\sigma} (l+r) \Ntil_{\alpha\beta\gamma} (l+r,l)
D^{\beta\tau} (l) = D^{\sigma\tau} (l) - D^{\sigma\tau} (l+r) \,.
\label{DND}
\end{eqnarray}
Similarly, replacing the fermion lines by $W^+$-lines in
\Eq{V-V(fig)}, we obtain another WT identity, which in the tree-level
reads
\begin{eqnarray}
c^W_{\lambda\rho\alpha\beta} (l+r+s,l+r) - 
c^W_{\lambda\rho\alpha\beta} (l+s,l) = 
\null - r^\gamma N_{\lambda\rho\alpha\beta\gamma} (l+r+s,l,-r) \,.
\label{C-C}
\end{eqnarray}

We can have an extra $Z$-boson present in all diagrams involved in
the identity of \Eq{D-D}, which will give us the relation
\begin{eqnarray} -e \left[
\left( \begin{picture}(50,30)(0,0)
\SetWidth{1.5}
\Photon(0,0)(25,0)24
\Text(10,-7)[]{$l$}
\Photon(25,0)(50,0)24
\Text(40,-7)[]{$l+t$}
\ZigZag(25,0)(25,25)25
\Vertex(25,0)5
\end{picture} \right)
- 
\left( \begin{picture}(50,30)(0,0)
\SetWidth{1.5}
\Photon(0,0)(25,0)24
\Text(10,-7)[]{$l-r$}
\Photon(25,0)(50,0)24
\Text(40,-7)[]{$l+s$}
\ZigZag(25,0)(25,25)25
\Vertex(25,0)5
\end{picture} \right) \right]
= r^\delta
\left( \begin{picture}(50,30)(0,0)
\SetWidth{1.5}
\Photon(0,0)(25,0)24
\Text(10,-7)[]{$l$}
\Photon(25,0)(50,0)24
\Text(40,-7)[]{$l+s$}
\Photon(25,0)(50,25)23
\Text(45,25)[r]{$r^\delta$}
\ZigZag(25,0)(0,25)25
\Vertex(25,0)5
\end{picture} \right) \,.
\end{eqnarray}
At the tree level, this implies
\begin{eqnarray}
\Ntil_{\alpha\beta\gamma}(l,l+r+s) - \Ntil_{\alpha\beta\gamma}(l-r,l+s) 
= \null - r^\delta R_{\alpha\beta\gamma\delta} \,.
\label{N-N}
\end{eqnarray}

We can also write down a similar WT identity involving the graviton,
viz., 
\begin{eqnarray} -e \left[
\left( \begin{picture}(50,30)(0,0)
\SetWidth{1.5}
\Photon(0,0)(25,0)24
\Text(10,-7)[]{$l$}
\Photon(25,0)(50,0)24
\Text(40,-7)[]{$l+r$}
\Graviton(25,0)(50,25)23
\ZigZag(25,0)(0,25)25
\Vertex(25,0)5
\end{picture} \right)
- 
\left( \begin{picture}(50,30)(0,0)
\SetWidth{1.5}
\Photon(0,0)(25,0)24
\Text(10,-7)[]{$l-r$}
\Photon(25,0)(50,0)24
\Text(40,-7)[]{$l$}
\Graviton(25,0)(50,25)23
\ZigZag(25,0)(0,25)25
\Vertex(25,0)5
\end{picture} \right) \right]
= k^\nu
\left( \begin{picture}(50,30)(0,0)
\SetWidth{1.5}
\Photon(0,0)(25,0)24
\Text(10,-7)[]{$l$}
\Photon(25,0)(50,0)24
\Text(40,-7)[]{$l$}
\Photon(25,0)(50,25)23
\Text(45,25)[r]{$r^\nu$}
\ZigZag(25,0)(0,25)25
\Graviton(25,0)(25,25)23
\Vertex(25,0)5
\end{picture} \right) \,.
\end{eqnarray}
The tree-level version of this identity reads
\begin{eqnarray}
N_{\lambda\rho\alpha\beta\mu}(l,l+r,r+s) - 
N_{\lambda\rho\alpha\beta\mu}(l-r,l,r+s) =
\null - r^\nu R_{\lambda\rho\alpha\beta\mu\nu} \,.
\label{N5-N5}
\end{eqnarray}

It is also instructive to see how some of these contractions behave
under the integration over the loop momentum.  For example, using
\Eq{DND}, we can write
\begin{eqnarray}
r^\gamma \lint l  && \hspace{-2em}
\Ntil_{\sigma\tau\delta} (l,l+r) D_W^{\alpha\sigma}(l+r)
\Ntil_{\alpha\beta\gamma} (l+r,l) D_W^{\beta\tau} (l)  \nonumber\\* 
&=& \lint l \Ntil_{\sigma\tau\delta} (l,l+r) \Big[
  D_W^{\sigma\tau} (l) - D_W^{\sigma\tau} (l+r) \Big] \nonumber\\ 
&=& \lint l \Big[ \Ntil_{\sigma\tau\delta} (l,l+r) -
  \Ntil_{\sigma\tau\delta} (l-r,l) \Big] D_W^{\sigma\tau} (l) \nonumber\\ 
&=& \null - r^\alpha R_{\sigma\tau\delta\alpha} \lint l D_W^{\sigma\tau}
(l) \,, 
\label{NDND}
\end{eqnarray}
where we have used \Eq{N-N} in arriving at the last step.  Through an
exactly similar kind of argument, we can prove the relation
\begin{eqnarray}
r^\gamma \lint l 
N_{\lambda\rho\sigma\tau\delta} (l,l+r,r+s) D_W^{\alpha\sigma}(l+r)
N_{\alpha\beta\gamma} (l+r,l,-r) D_W^{\beta\tau} (l)  
\nonumber\\*  
= \null - r^\alpha R_{\lambda\rho\sigma\tau\delta\alpha} \lint l
D_W^{\sigma\tau} (l) \,.
\label{N5DND}
\end{eqnarray}
%

\section{Some other relations between various couplings}\label{s:or}
In Appendix \ref{s:WT}, we considered the contraction of various
vertices with the photon momentum.  In this Appendix, we are
considering contractions of a more general kind, in particular
relation involving the graviton momentum.

From the definitions of these vertices in \Eqs{N}{R4}, it is easy to
see that one can write a relation between the cubic and the quartic
gauge couplings:
\begin{eqnarray}
\Ntil_{\alpha\beta\gamma}(a,b) 
= a^\delta R_{\gamma\alpha\beta\delta} + b^\delta
R_{\beta\gamma\alpha\delta} \,.
\label{NR}
\end{eqnarray}
It is easy to see that \Eq{N-N} can be derived from this relation by
making use of the identity
\begin{eqnarray}
R_{\alpha\beta\gamma\delta} + R_{\beta\gamma\alpha\delta} +
R_{\gamma\alpha\beta\delta} = 0 \,,
\label{RRR}
\end{eqnarray}
which follows trivially from the expression for the quartic gauge
coupling in \Eq{R4}.  The following identities, used at different
stages of the calculation, can also be derived from \Eq{NR}:
\begin{eqnarray}
\Ntil_{\alpha\beta\gamma}(a,b) - \Ntil_{\alpha\beta\gamma}(a,b-r) 
&=& r^\delta R_{\beta\gamma\alpha\delta} \,, 
\label{Nab-Nab'}\\ 
\Ntil_{\alpha\beta\gamma}(a,b) - \Ntil_{\alpha\beta\gamma}(a-r,b) 
&=& r^\delta R_{\gamma\alpha\beta\delta} \,.
\label{Nab-Na'b}
\end{eqnarray}

Another important relation involves the contraction of the coupling
$N_{\lambda\rho\alpha\beta\gamma}$ with the graviton momentum.  From
the definition in \Eq{N5}, it easily follows that 
\begin{eqnarray}
q^\lambda N_{\lambda\rho\alpha\beta\gamma} (a,a+b,b+q)
&=& \Big( q_\rho - a_\rho + b_\rho \Big) \Ntil_{\alpha\beta\gamma}
(a,a+b) 
\nonumber\\* && 
+ a_\rho \Ntil_{\alpha\beta\gamma} (a-q,a+b-q)
- b_\rho \Ntil_{\alpha\beta\gamma} (a,a+b+q)
\nonumber\\* && 
- q^\delta \Big( \eta_{\rho\alpha} \Ntil_{\delta\beta\gamma} (a-q,a+b)
+ \eta_{\rho\beta} \Ntil_{\alpha\delta\gamma} (a,a+b+q)
\nonumber\\* && \hspace{2cm}
+ \eta_{\rho\gamma} \Ntil_{\alpha\beta\delta} (a,a+b) \Big) . \qquad
\label{q.N5NN}
\end{eqnarray}
Sometimes alternative forms of this identity is more useful, such as 
\begin{eqnarray}
q^\lambda N_{\lambda\rho\alpha\beta\gamma} (a,a+b,b+q)
&=& q_\rho \Ntil_{\alpha\beta\gamma} (a,a+b)
+ q^\delta \Big( a_\rho R_{\alpha\beta\gamma\delta} 
- b_\rho R_{\beta\gamma\alpha\delta} \Big) 
\nonumber\\* && 
- q^\delta \Big( \eta_{\rho\alpha} \Ntil_{\delta\beta\gamma} (a-q,a+b)
+ \eta_{\rho\beta} \Ntil_{\alpha\delta\gamma} (a,a+b+q)
\nonumber\\* && \hspace{2cm}
+ \eta_{\rho\gamma} \Ntil_{\alpha\beta\delta} (a,a+b) \Big) ,
\label{q.N5RR}
\end{eqnarray}
or
\begin{eqnarray}
q^\lambda N_{\lambda\rho\alpha\beta\gamma} (a,a+b,b+q)
&=& (b_\rho + q_\rho) \Ntil_{\alpha\beta\gamma} (a,a+b)
- b_\rho \Ntil_{\alpha\beta\gamma} (a,a+b+q)
+ q^\delta a_\rho R_{\alpha\beta\gamma\delta} 
\nonumber\\* && 
- q^\delta \Big( \eta_{\rho\alpha} \Ntil_{\delta\beta\gamma} (a-q,a+b)
+ \eta_{\rho\beta} \Ntil_{\alpha\delta\gamma} (a,a+b+q)
\nonumber\\* && \hspace{2cm}
+ \eta_{\rho\gamma} \Ntil_{\alpha\beta\delta} (a,a+b) \Big), 
\label{q.N5RN}
\end{eqnarray}
which can be derived from \Eq{q.N5NN} by using the identities of
\Eqs{Nab-Nab'}{Nab-Na'b}.


\begin{thebibliography}{[99]}

\bibitem{Yang:1950rg}
  C.~N.~Yang,
  Phys.\ Rev.\  {\bf 77}, 242 (1950).

\bibitem{Bjerrum-Bohr:2002kt}
  N.~E.~J.~Bjerrum-Bohr, J.~F.~Donoghue and B.~R.~Holstein,
  Phys.\ Rev.\ D {\bf 67}, 084033 (2003)
  [Erratum-ibid.\ D {\bf 71}, 069903 (2005)]
  [arXiv:hep-th/0211072].


\bibitem{Bjerrum-Bohr:2002ks}
  N.~E.~J.~Bjerrum-Bohr, J.~F.~Donoghue and B.~R.~Holstein,
  Phys.\ Rev.\ D {\bf 68}, 084005 (2003)
  [Erratum-ibid.\ D {\bf 71}, 069904 (2005)]
  [arXiv:hep-th/0211071].


\bibitem{Weinberg:1978kz}
  S.~Weinberg,
  PhysicaA {\bf 96}, 327 (1979).

\bibitem{Donoghue:1994dn}
  J.~F.~Donoghue,
  Phys.\ Rev.\ D {\bf 50}, 3874 (1994)
  [arXiv:gr-qc/9405057].

\bibitem{Burgess:2003jk}
  C.~P.~Burgess,
  Living Rev.\ Rel.\  {\bf 7}, 5 (2004)
  [arXiv:gr-qc/0311082].


\bibitem{Choi:1994ax}
  S.~Y.~Choi, J.~S.~Shim and H.~S.~Song,
  Phys.\ Rev.\ D {\bf 51} (1995) 2751
  [arXiv:hep-th/9411092].

\bibitem{Shim:1995ap}
  J.~S.~Shim and H.~S.~Song,
  Phys.\ Rev.\ D {\bf 53} (1996) 1005
  [arXiv:hep-th/9510024].


\bibitem{Nieves:1998xz}
  J.~F.~Nieves and P.~B.~Pal,
  Phys.\ Rev.\ D {\bf 58} (1998) 096005
  [arXiv:hep-ph/9805291].

\bibitem{Nieves:1999rt}
  J.~F.~Nieves and P.~B.~Pal,
  Mod.\ Phys.\ Lett.\ A {\bf 14} (1999) 1199
  [arXiv:gr-qc/9906006].

\bibitem{Nieves:2000dc}
  J.~F.~Nieves and P.~B.~Pal,
  Phys.\ Rev.\ D {\bf 63} (2001) 076003
  [arXiv:hep-ph/0006317].

\bibitem{scadron} Michael D. Scadron, Advanced Quantum Theory,
Springer-Verlag (New York, 1979).

\end{thebibliography}
\end{document}